\documentclass[a4paper]{article}
\usepackage{graphicx}  
\usepackage{color}     
\usepackage{comment}
\usepackage{color}
\usepackage{lscape}
\usepackage{amssymb}
\usepackage{amsmath}
\usepackage{pstricks}
\usepackage{longtable}
\usepackage{feynmp}
\usepackage{breqn}

\date{\today}
 \normalsize

\newcommand{\gtr}{\mbox{${\sf SU(3)_C \times SU(3)_L \times SU(3)_R \times
{Z}_3}$}}  
\newcommand{\bmat}{\left(\begin{array}}
\newcommand{\emat}{\end{array}\right)}
\newcommand{\be}{\begin{equation}}
\newcommand{\ee}{\end{equation}}
\newcommand{\bea}{\begin{eqnarray}}
\newcommand{\eea}{\end{eqnarray}}

\def\Lag{{\mathcal{L}}}

\def\gtwid{\mathrel{\raise.3ex\hbox{$>$\kern-.75em\lower1ex\hbox{$\sim$}}}}
\def\ltwid{\mathrel{\raise.3ex\hbox{$<$\kern-.75em\lower1ex\hbox{$\sim$}}}}
\def\gev{{\rm \, Ge\kern-0.125em V}}
\def\tev{{\rm \, Te\kern-0.125em V}}

\def    \be            {\begin{equation}}
\def    \ee            {\end{equation}}
\def    \bea           {\begin{eqnarray}}
\def    \eea           {\end{eqnarray}}
\def\eps{\epsilon}
\def\a{\alpha}
\def\b{\beta}
\def\g{\gamma}
\def\d{\delta}

\def\Om{\Omega}

\def\sig{\sigma}

\def\lam{\lambda}

\def\m{\mu}
\def\nn{\nonumber}
\def\d{\delta}

\def\t{\theta}

\begin{document}
\renewcommand{\thefootnote}{\fnsymbol{footnote}}

\vspace{.3cm}

\title{\Large\bf Six Higgs Doublets Model for Dark Matter}

\author
{ \hspace{-3.cm} \it \bf  Mohammad Alakhras$^{1}$\thanks{m.alakhras@damasuniv.edu.sy},  Nidal
Chamoun$^{2,4}$\thanks{nchamoun@th.physik.uni-bonn.de}, Xue-Lei Chen$^{3}$\thanks{xuelei@cosmology.bao.ac.cn},
 Chao-Shang Huang$^{4}$\thanks{csh@itp.ac.cn} and Chun Liu$^{4}$\thanks{liuc@mail.itp.ac.cn}
 \\\hspace{-3.cm}
\footnotesize$^1$ Department of Physics, Faculty of Sciences, Damascus University, Damascus,
Syria.\\\hspace{-3.cm}
\footnotesize$^2$  Physics Department, HIAST, P.O.Box 31983, Damascus,
Syria.\\\hspace{-3.cm}
\footnotesize$^3$ Key Laboratory of Computational Astrophysics,
National Astronomical Observatories, CAS, Beijing 100012, China. \\\hspace{-3cm}
 \footnotesize$^4$ CAS Key Laboratory of Theoretical Physics, Institute of Theoretical Physics, CAS,
P.O. Box 2735, Beijing 100190, China.
}
%

\date{\today}

\maketitle
\begin{abstract}
We consider an extension of the Higgs sector in the standard model
(SM) with six Higgs doublets. The gauge couplings are unified
without supersymmetry in this model. The lightest of the extra
Higgs particles, being stabilized by a discrete symmetry imposed
from the outset, presents a plausible candidate for dark matter. For a specific acceptable benchmark point, we show that the model is viable regarding the
constraints of relic density, direct detection and invisible Higgs decay. We comment on the mean free path of the dark matter candidate.
\end{abstract}

\section{Introduction}

As is well known, baryonic matter constitutes only 4-5\% of the
total cosmic energy density. About 20\% of the Universe is made up
of dark matter (DM), and the remaining part is dark energy
\cite{planck2016}. Although the first evidence of DM was found
seven decades ago, we still do not know its composition, and
finding clues for its nature is one of the most pressing issues in
physics. Since none of the particles in the Standard Model (SM) can be a viable
candidate of the major part of the DM, a successful extension of
the SM of particle physics is likely to address this problem.
Interestingly, Weakly Interacting Massive Particles (WIMPs) can be
 very good candidates of the cold DM (CDM), since their relic
abundances agree naturally with astrophysical observations. Many
particle physics models were proposed involving new, other than
SM, particles having masses at the weak scale, and which couple
weakly to the SM particles.

In supersymmetric (SUSY) models with $R$-parity
conservation, the lightest SUSY particle (LSP) is
stable, and is a DM candidate. The most popular LSP is the
neutralino, a superposition of photino, zino and the two
Higgsinos. The experimental limit on the mass of the lightest
neutralino, $\tilde{\chi}^0_1$, due to the negative searches at
accelerators such as LHC, is $m_{\tilde{\chi}^0_1} > 46$ GeV \cite{plb}. Other possible SUSY DM
candidates include gravitino and sneutrino.

Moreover, SUSY allows also for coupling unification of the strong, weak, and electromagnetic
interactions, and pushes the unification
scale in $SU(5)$ models from $10^{15}$ GeV in its absence up to
around $10^{16}$ GeV. This likely makes the $SU(5)$ GUT safe from rapid
proton decay \cite{dimopoulos81}\footnote{The minimal SUSY SU(5) has been already
excluded by the limit $(6.7 \times 10^{32} \mbox{ y})$ of the proton
lifetime from Super Kamiokande \cite{hep-ph/0108104}. Nevertheless, there are a number of ways in
which this limit can be satisfied (for a review, see
\cite{hep-ph/0601023})}.

Although well motivated, SUSY models are theoretically quite
complicated. Moreover, the negative searches for its simple versions at the LHC impose stiff constraints on SUSY models. In fact, there are other ways to achieve gauge
coupling unification. Upon closer examination of SUSY models, the
modification on the running of the gauge coupling is due entirely
to the extension of the Higgs sector, which includes a second Higgs
doublet and the corresponding superpartners. It has been known
that the unification of the gauge couplings can also be achieved
without SUSY, if the SM has an extended Higgs sector with
six Higgs doublets at the weak scale \cite{will3}. Furthermore, the problem of rapid proton decay can be tamed
in non-SUSY extended
Higgs models, such as the trinification model \cite{babu86}, where the GUT gauge group
is not a simple group but the product group $\gtr$,
and where the proton life time can be above the experimental limits
\cite{will1}.

For the
non-SUSY gauge unification model, a very interesting question arises as to whether or not this model can provide an explanation for the DM problem.
Various non-SUSY models for DM were proposed ranging
from axions \cite{turner90}, sterile neutrinos \cite{lee77}, to lightest
Kaluza-Klein excitations in universal extra dimensions models
\cite{antoniadis90}, and/or branons where the fluctuations of the
branes in string theory can be made into suitable CDM
candidates \cite{cembranos03}. SM-singlet extensions \cite{burgess01,nasri} are popular DM models. The additional SM-singlet scalars have no direct coupling to gauge fields, and so no annihilation into gauge bosons unless via intermediate Higgs. Thus these models are mainly Higgs-portal models. Since the SM Higgs sector is based on doublets, one can consider another way going beyond SM by extending the Higgs sector. Here, the DM can be accommodated considering certain combinations of the additional doublets, and the DM direct annihilations into gauge vector bosons exist leading to known signatures compared to singlet extensions. The inert Doublet Model (IDM) \cite{ma06,barbieri06}, where one adds another `inert' doublet to the SM Higgs doublet, is a starting point for many extensions, can accommodate axions while still being testable experimentally \cite{Queiroz},  and has been suggested
as a simple and yet rich model for DM \cite{honorez07,dolle09}.

In this work we study the DM problem within the
framework of multi-doublet scalars without SUSY. In particular, we
extend the SM Higgs sector to contain six Higgs doublets, and
consider the possibility of one of the `additional' Higgs
particles being the DM. We show that if one imposes a
discrete symmetry, then one of the Higgs particles can be
stabilized, thus providing a possible candidate of DM. Thus,
extending the Higgs sector in a non-SUSY context helps
to realize two objectives: gauge unification and accounting for
DM. More specifically, we assume that two of the Higgs doublets ($H_u, H_d$) get
vacuum expectation values (VEV), to break the electroweak symmetry
and generate masses for the quarks and charged leptons. This part of the
model is similar to the two Higgs doublet model (2HDM) \cite{LGW,haber79} and
is phenomenologically viable \cite{Branco2011,Huang2HDM}. For simplicity, we call hereafter this part the ``active doublets" whereas we call the remaining part the ``inert doublets" . We denote the lightest
neutral component of the additional Higgs particles (LAH), in inert doublets, by
$\psi$ which, due to the discrete symmetry imposed, can not decay
to SM Higgses ($H_u,H_d$), nor can it have Yukawa-type
interactions.

Actually, DM with one Higgs `active' doublet plus two `inert' doublets (denoted 1+2), considered the simplest extension to the DM with (1+1) IDM,  was studied in \cite{king1,king2} and shown to have a richer phenomenology than the IDM or the (2+0) 2HDM. Motivated by SUSY which requires an even number of doublets, the authors of \cite{king1} extended the (1+2) model into a (2+4) model with two Higgs active doublets and four inert doublets, and computed the mass spectrum of the model for a special form of the Lagrangian $\Lag_0$.

Besides this heuristic argument in going from the IDM (1+1) to the (2+4) passing through (1+2), we note moreover that although the LHC Higgs is SM-like, however many approaches posses a decoupling limit behaving like the SM, and the 2HDM provides a framework for studying the decoupling limit and possible departures from SM-like Higgs behavior \cite{Haber_2013talk}.

We thus take the same (2+4) extension motivated, in addition, by the unification of coupling constants plus accounting for DM, and we build it with the most general form of the Lagrangian, containing $\Lag_0$ of \cite{king1}, allowed by symmetry. We did not consider the (1+5) extension as it resembles the (1+1) IDM except for having more particles some of which may be close in mass to the DM mass, and thus more possible co-annihilation channels exist. The extension (2+4) is far richer since, compared to the SM, we changed both the inert and the active sectors. In fact, the 2HDM is chosen rather than the SM because it is under scrutiny at present in view of the LHC results \cite{das}, and it is a simple extension of the SM.

For the phenomenological analysis, and since we focus in this work on the model building and its feasibility for providing a DM candidate, we took a special benchmark point in the parameter space, called ``symmetric choice", which mimics the case rarely studied of the (2+1) model.  In this simplest realization, the model will have three free
parameters: the mass of the new LAH $m_\psi$, a dimensionless
self-coupling $\lambda_S$ and a dimensionless coupling $\lambda$
to the SM Higgs ($H$). If the LAH is the DM, then $\lambda$ and
$m_\psi$ are related by requiring it to give the cosmological DM
abundance. In addition to the Relic Density (RD) constraint, we imposed also the constraints originating from the Invisible Higgs Decay (IHD) and the negative searches of the Direct Detection (DD) experiments related to the DM elastic scattering with atomic nuclei. We find that the mass of the LAH should be around the Higgs resonance
($m_\psi \approx \frac{1}{2} m_H$) in order to have a perturbative coupling ($\lambda
\leq 1$), to account for the DM abundance and to respect the IHD and DD constraints.
As a matter of fact, the authors of \cite{grzadowski} studied the extension (2+1) where an inert doublet was added to the 2HDM, and explored the $CP$ violation effects in it. However, no IHD constraints were considered there, and no phenomenology relating the DM mass versus  the coupling $\lambda$ between the active and inert Higgs doublets was presented. Moreover, we have taken into account the new updated experimental constraints, and in particular we have imposed the newly discovered Higgs mass value.

In this work we have computed the scattering$/$decay amplitudes using the Mathematica packages
(SARAH, FeyArts, FormCalc) \cite{sarah}, and then obtained the thermal relic
abundance by using approximate formulae\cite{kolbturner,sredinski}. This enables us to
see the details of the various processes, and is sufficient
for checking the basics of the model. The more sophisticated package
such as micrOMEGAs \cite{micromegas}, while more complete and accurate,
hides much physical details under its hood. However, we checked
that for the chosen benchmark point of parameter space considered
here, the micrOMEGAs package gives results in line with what we get here.

The paper is organized as follows. In section 2, we review the gauge coupling unification with six Higgs doublets in subsection (2.1), and build our (2+4) model presenting the most general form of the Lagrangian in subsection (2.2). We implement sufficient conditions for the stability of the potential in subsection (2.3). In subsection (2.4) we present the sector representing the active Higgs doublets, whereas subsection (2.5) presents the inert sector.  Section 3  is devoted to our ``symmetric choice'' benchmark point stated in subsection (3.3). In subsections (3.1, 3.2) we determine the parameter space point in the active and inert sectors respectively.
We present the phenomenological analysis in section 4. Subsection (4.1) presents the RD  computations, whereas subsection (4.2) presents the DD constraints. Subsection (4.3) presents the IHD constraints, and finally we comment on the DM self-elastic scattering in subsection (4.4). Section (5) presents our conclusions about the viability of the model in the ``symmetric choice'' benchmark point. There are few appendices where several technical formulae are listed.

\section{Building the Model}

\subsection{Motivation--Unification of gauge couplings}

In order to achieve gauge unification in the model, let
us review how this is done in the SUSY $SU(5)$ model.
The fermion superpartners in the SUSY model do not affect the
relative evolution of the
gauge couplings, since they fall in complete representations of
the $SU(5)$ group, while the boson superpartners change the unification
scale. Crucially, it is the existence of second Higgs doublets, as
well as the Higgs superpartners which modifies the relative
evolution of the gauge couplings,
because the Higgs doublets do not form a complete
representation of the $SU(5)$ group.

 The renormalization group equation for the gauge
couplings $\a_k = \frac{g_k^2}{4 \pi}$, reads:
\bea
\frac{1}{\alpha_k(\mu)}-\frac{1}{\alpha_k(\mu')}&=&\frac{b_k}{2\pi}
\ln\left(\frac{\mu'}{\mu}\right)\;,\label{rng}
\eea
 where the $b_k$'s are the beta-functions given by
 \bea \label{beta}
 \frac{d g_k}{d \log{\mu}}\equiv\b_k(g_k)\equiv \frac{\beta_{g_k}^{(1)}}{(4 \pi)^2} + \frac{\beta_{g_k}^{(2)}}{(4 \pi)^4} &=& b_k \frac{g_k^3}{(4 \pi)^2} + \sum_{\ell=1}^{\ell =3} b_{k \ell}\frac{g_k^3 g_\ell^2}{(4 \pi)^4} + \frac{g_k^3}{(4 \pi)^4}
 \mbox{Tr}\left( C^u_\ell Y_u Y_u^\dagger + C^d_\ell Y_d Y_d^\dagger + C^e_\ell Y_e Y_e^\dagger\right) \nn \\
 \eea
In appendix \ref{rge}, we find the following expressions:
\bea
\beta_{g_1}^{(1)} & = &
\frac{23}{5} g_{1}^{3} \label{rge_g1^1} \\
\beta_{g_1}^{(2)} & = &
\frac{1}{50} g_{1}^{3} \Big(244 g_{1}^{2}  -25 \mbox{Tr}\Big({Y_d  Y_{d}^{\dagger}}\Big)  + 360 g_{2}^{2}  + 440 g_{3}^{2}  -75 \mbox{Tr}\Big({Y_e  Y_{e}^{\dagger}}\Big)  -85 \mbox{Tr}\Big({Y_u  Y_{u}^{\dagger}}\Big) \Big) \label{rge_g1^2}\\
\beta_{g_2}^{(1)} & = &
-\frac{7}{3} g_{2}^{3} \label{rge_g2^1}\\
\beta_{g_2}^{(2)} & = &
\frac{1}{30} g_{2}^{3} \Big(-15 \mbox{Tr}\Big({Y_e  Y_{e}^{\dagger}}\Big)  + 360 g_{3}^{2}  -45 \mbox{Tr}\Big({Y_d  Y_{d}^{\dagger}}\Big)  -45 \mbox{Tr}\Big({Y_u  Y_{u}^{\dagger}}\Big)  + 500 g_{2}^{2}  + 72 g_{1}^{2} \Big) \label{rge_g2^2}\\
\beta_{g_3}^{(1)} & = &
-7 g_{3}^{3} \label{rge_g3^1}\\
\beta_{g_3}^{(2)} & = &
-\frac{1}{10} g_{3}^{3} \Big(-11 g_{1}^{2}  + 20 \mbox{Tr}\Big({Y_d  Y_{d}^{\dagger}}\Big)  + 20 \mbox{Tr}\Big({Y_u  Y_{u}^{\dagger}}\Big)  + 260 g_{3}^{2}  -45 g_{2}^{2} \Big) \label{rge_g3^2}
\eea

\begin{figure}[hbtp]
\centering
\begin{minipage}[l]{0.5\textwidth}
\centerline{\includegraphics[width= \linewidth]{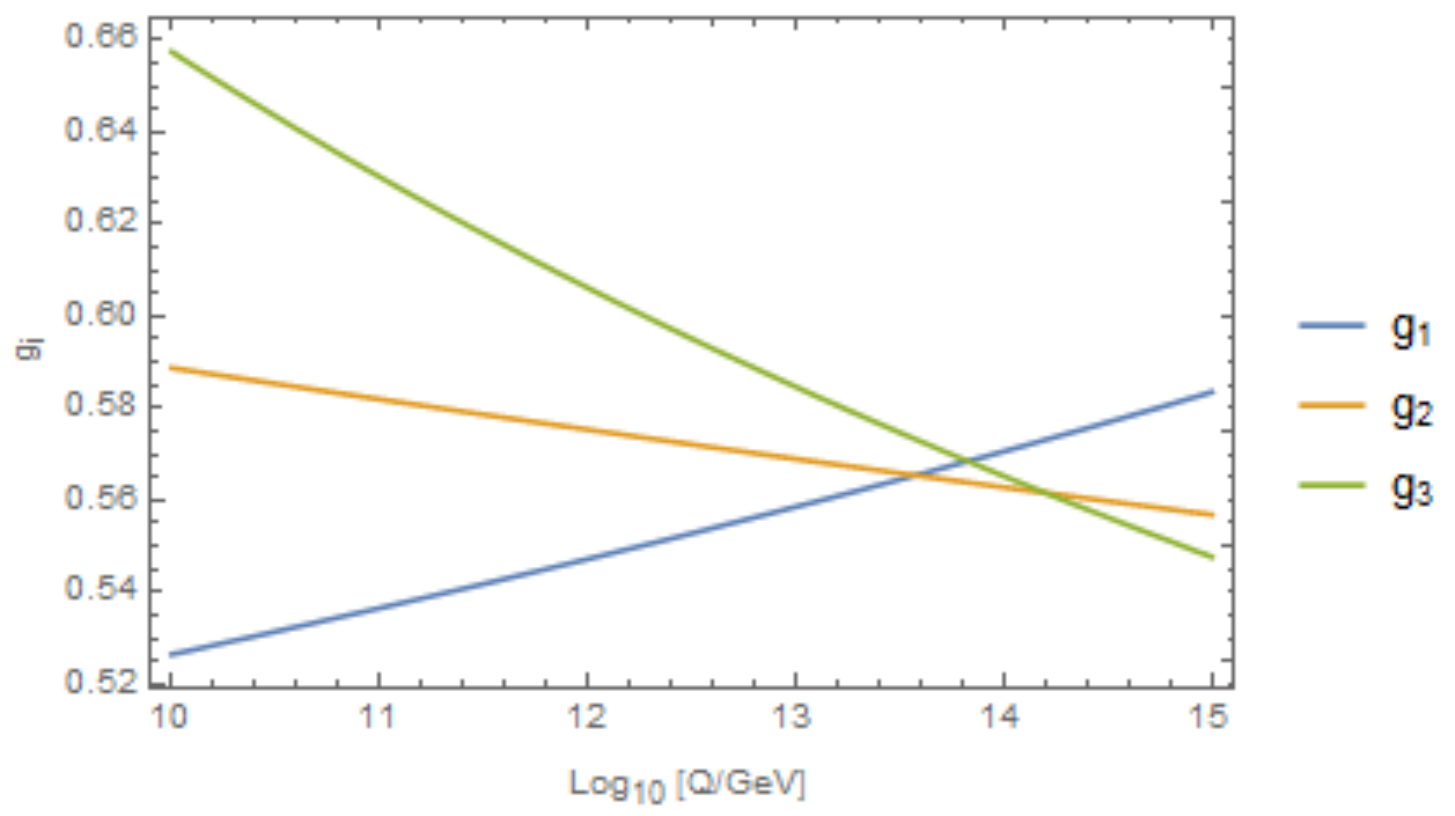}}
\end{minipage}%
\begin{minipage}[r]{0.5\textwidth}
\centerline{\includegraphics[width= \linewidth]{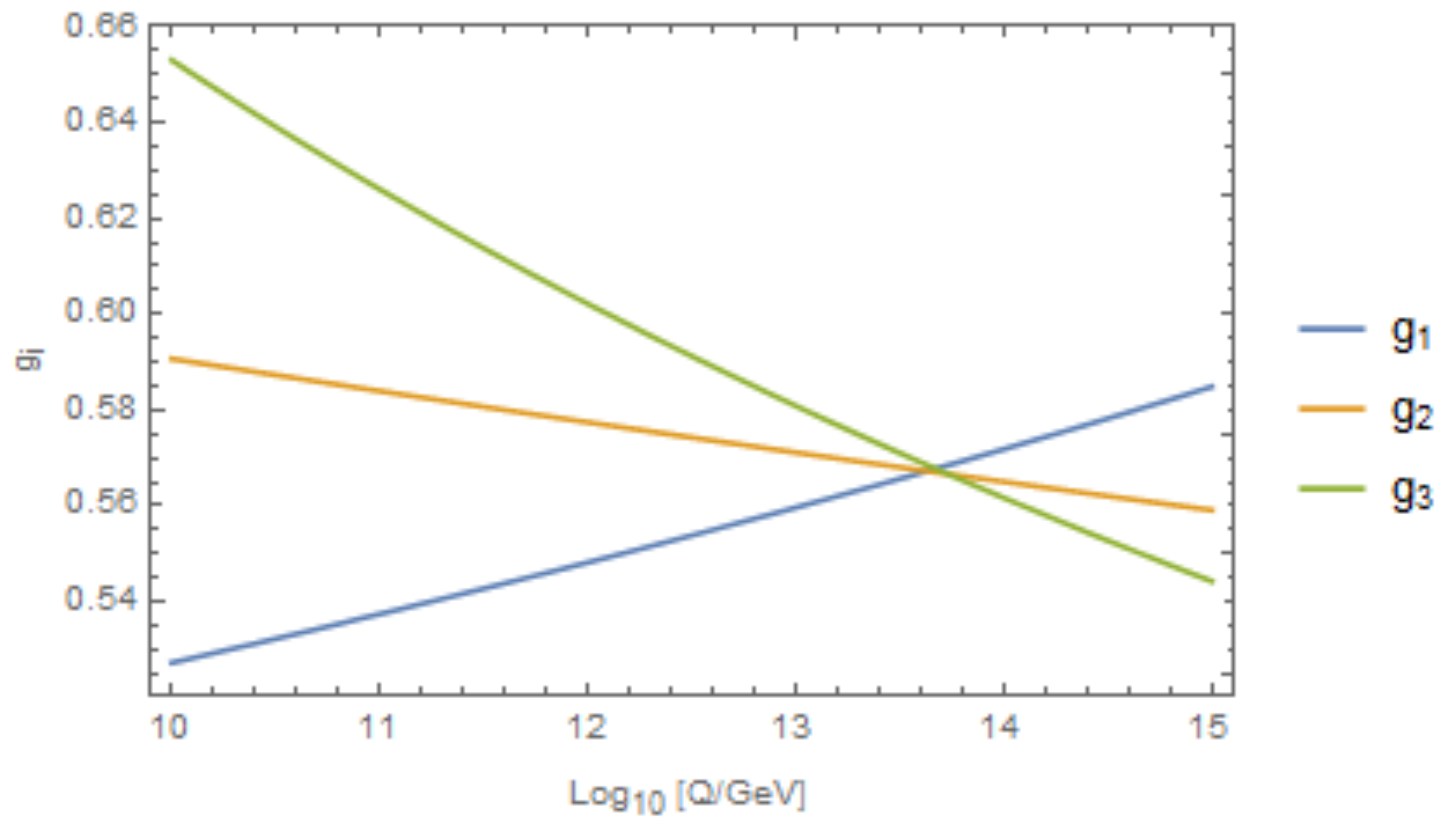}}
\end{minipage}
\vspace{0.5cm}
\caption{{\footnotesize  Unification of gauge  couplings in SM with $n_H=6$ Higgs doublets at 1-loop (left) and 2-loop (right) runnings. }}
\label{fig:unification}
\end{figure}

We plot in Fig.~\ref{fig:unification} the 1--loop (left) and 2--loop (right) running couplings for the particle content of the SM but with $6$ Higgs doublets, and with the inputs $\alpha_3(M_Z) =
0.117$, $\alpha_2(M_Z) = (\sqrt 2/\pi)G_FM_W^2 = 0.034$ and
$\alpha_1(M_Z) = (5/3)\alpha_2(M_Z)\tan^2\theta_W=0.017$. We limited the Yukawas couplings to the third family and did not consider the ``higher order'' effect of the masses' running, so took, say, $\mbox{Tr} Y_u Y_u^\dagger = 2\frac{m^2_t}{v^2} \approx 2 (\frac{175}{246})^2$. We see the 2-loop unification is better satisfied than that of the 1--loop, and that the unification scale occurs at around $M_U \sim 10^{13.7} \gev$. This low
unification scale yields rapid proton decay in conventional
$SU(5)$ theory. In order to avoid this problem, we consider,
as Ref.~\cite{will1,will3}, the possibility that the unified
theory is not $SU(5)$ but a trinified model based on $\gtr$, with
a Higgs sector containing six \footnote{In fact, one can get gauge
unification with either five or six Higgs doublets. However, to keep with
the model triplication paradigm, we assume six such Higgses.} Higgs
fields in the $27$-dim representation of $(SU(3))^3$. Those
fields, upon symmetry breaking
$$(SU(3))^3 \rightarrow SU(3)_c \times SU(2)_L \times U(1)_Y,$$
leads to the SM with six Higgs doublets, of which one linear
combination acquires a mass of order of the unification scale
while the remaining five doublets may have the mass of order of
the weak scale \cite{will1}\footnote{The work of \cite{Dias} also considered SM with 6 Higgs doublets and showed that with an invisible axion it is possible to obtain proton stability and coupling unification at an energy near the Peccei-Quinn scale.}.

\subsection{The Lagrangian}

We denote the six Higgs fields of hyper charge $Y=1$ by $H_I (I=u,d),H_\a
(\a=1,\dots,4)$. The electroweak symmetry is broken to
$U(1)_{em}$ when any of the six Higgs doublets acquires an
electroweak scale VEV. For simplicity, we follow \cite{will1} and
assume that two such Higgses ($H_I (I=u,d)$) get VEVs: $ H_u$ to give masses
 to the u-type quarks and $H_d$ to give masses
 to the d-type quarks and to the charged leptons. This parallels the
 structure of the known two-Higgs doublet model (type II). In order to
 naturally suppress flavor changing neutral currents, i.e., to
 have natural flavor conservation (NFC), one imposes some discrete
 symmetry and a simplest choice is a $Z_2$ symmetry:
 \bea H_u \to -H_u,~~~~ u_R\to -u_R \eea
and all the other fields unchanged under the $Z_2$ transformation.
To keep at least one of the other Higgses stable, we impose another
discrete symmetry $Z_2^\prime$ under which
$H_\alpha \to -H_\alpha (\a=1,\dots,4)$ while all other fields are unchanged.

The general form of the renormalizable $SU(2)_L\times U(1)_Y$ gauge invariant Lagrangian which
is invariant under the discrete symmetry $Z_2\times Z_2^\prime$ is
given by:
\be \label{Ldef}
\Lag = \Lag_{\mbox{Kin}} - V(H)- V_{\mbox{Yuk}}
\ee
where:
\bea
\Lag_{\mbox{Kin}} &= &\left( D_\mu H_I \right)^\dagger \left( D_\mu H_I \right) + \left( D_\mu H_\a \right)^\dagger \left( D_\mu H_\a \right) \\ V(H) &=& V_{\mbox{\tiny Active}} + V_{\mbox{\tiny Inert}} + V_{\mbox{\tiny Coupled}} \label{Vdef}  \\
V_{\mbox{\tiny Active}} &= &\mu_{II} H_I^\dagger H_I + \lam_{IIJJ} H_I^\dagger H_I H_J^\dagger H_J + \Big( \eta_{IJJI} H_I^\dagger H_J H_J^\dagger H_I  \label{VActive}\\
&+ &\frac{1}{2}\left( \xi_{IJIJ} H_I^\dagger H_J H_I^\dagger H_J + h.c. \right) \Big)_{I\neq J} \\
V_{\mbox{\tiny Inert}} &= & \mu_{\a \b} H_\a ^\dagger H_\b +  \lam_{\a \b \g \d} H_\a ^\dagger H_\b H_\g ^\dagger H_\d + h.c.  \label{VInert}\\
V_{\mbox{\tiny Coupled}} &= & \lam_{II\a \b} H_I^\dagger H_I H_\a ^\dagger H_\b + \eta_{I\a \b I} H_I^\dagger H_\a H_\b ^\dagger H_I + \frac{1}{2} \left( \xi_{I\a I\b} H_I^\dagger H_\a H_I^\dagger H_\b + h.c. \right) \label{VCoupled},\\
V_{\mbox{Yuk}} &= &Y_l L^\dagger H_d e_R + Y_d Q H_d d_R + Y_u Q \tilde{H}_u u_R \label{Vyukawa}
\eea
with $\tilde{H}_u=i\tau_2 H_u^*$, the covariant derivative
$D_\mu = \partial _\mu + i g W^{I\nu} \frac{\sigma^I}{2}
 g_{\mu\nu} + i g^\prime B^\nu g_{\mu\nu}$,
and the $W$ and $B$ are the gauge bosons for $SU(2)_L \times
U(1)_Y$, and where the Latin indices ($I,J$)  run over the set $\{u,d\}$, whereas the Greek indices ($\a, \b$) span the set $\{1,\ldots,4\}$.

Before proceeding some remarks are in place. First, from the
Lagrangian (\ref{Ldef}), the
lightest among the additional Higgses (LAH), say, the neutral
component $\psi$ of $(H_3)^t \equiv (H_\a^+,
\phi_3 + i \sigma_3)$, is stable since it can not decay to other
 particles, and thus is a possible candidate for the DM.  Our
 restriction that $H_I$'s only have a non-zero electroweak VEV means
 that the $Z_2^\prime$ symmetry is unbroken after the electroweak symmetry breaking
. If one, instead of $Z_2^\prime$,
 imposes a symmetry $R$ of a reflection on each additional
 Higgs doublet, i.e. $R=\prod_{\beta = 1}^{\beta = 4} Z_2^{(\beta)}$ where under
 each $Z_2^{(\beta)}$ we have $H_\beta\to - H_\beta$ whereas
 all the other fields are unchanged, then we have
 ~\cite{weinb,ST}:
 \begin{equation} \label{V1def} V(H) = \Sigma_i[\mu_i H_i^\dagger H_i +
 \lam_{ii}(H_i^\dagger H_i)^2]+ \Sigma_{i<j}[ \lam_{ij} H_i^\dagger H_i H_j^\dagger
 H_j+  \eta_{ij}H_i^\dagger H_j H_j^\dagger
 H_i + (\xi_{ij}H_i^\dagger H_j H_i^\dagger
 H_j + h.c.)]
\end{equation} where $i \in \{u,d,1,\ldots,4\}$.
In this situation, the lighter of neutral Higgs bosons for each
additional doublet is stable and can be a candidate for
the DM. For simplicity we do not pursue this case in the paper.
Next, if all coupling constants are real the potential is
CP-invariant. There is also no spontaneous CP breaking because we
have assumed that only two doublets of all the six doublets can
get non-zero vevs and there are no linear terms of the form
$H_u^+H_d$ due to the $Z_2$ symmetry. We do not discuss CP
violation in the paper so that we assume all couplings are real
hereafter. Third, as it is obvious from the potential (\ref{Vdef})
(as well as (\ref{V1def})), we do not have a Peccei-Quinn $U(1)$
global symmetry, and thus axions at the elctro-weak scale do not
arise when the two Higgses take a vev.

\subsection{Stability}

We do not address here the question of whether the broken vacuum represents a global minimum of the potential, where, if false, it is necessary to compare the tunneling time for the metastable vacuum with the age of the universe, but rather we ensure it is a local minima by choosing to scan over the physical masses with positive values, whence the local minima conditions of positive eigenvalues of the mass matrix are satisfied automatically. This results from expanding the potential $V(\psi_i)$ around its extremum $\psi_i=v_i$: \bea V(\psi_i) &=& V(v_i) + \partial_i V (\psi_i-v_i) + \frac{1}{2} (\psi_i-v_i) \partial_{ij}V(\psi_j-v_j),\eea then the tadpole conditions ensures the vanishing of the linear term, whereas the positivity of the matrix $\partial_{ij}V$ (i.e. the positivity of its eigenvalues) ensures being at a local minimum. The eigenvalues of $\partial_{ij}V$ are nothing but the ``masses'' of the fields $\psi_i$.

Furthermore, one can impose sufficient conditions for the boundedness of the potential which we do now. We divided the potential into three parts ($V_{\mbox{\tiny Active}}, V_{\mbox{\tiny Inert}}$ and $V_{\mbox{\tiny Coupled}}$), so by requiring each part to be positive, and so bounded from below, we reach sufficient conditions, which may not be necessary ones, for the  potential boundedness from below. By Caushy-Schwartz inequality, we can set for any two complex vectors one angle $\t(1,2)$ and one phase $\phi(1,2)$ defined as follows.
\bea V_1^\dagger . V_2 = \mid V_1^\dagger . V_2 \mid e^{i \phi(1,2)}=\parallel V_1 \parallel   . \parallel V_2 \parallel  \cos\t(1,2) e^{i \phi(1,2)}, & \mbox{where} & \t \in [0,\pi/2], \phi \in [-\pi, \pi]
\eea

Restricting to fourth order terms, in order to study the large  field limit,
and assuming all the couplings are real, so $\xi_{udud} \equiv \xi^*_{dudu} =\xi_{dudu}$
\bea
V_{\mbox{\tiny Active}} &\supseteq& \lam_{uuuu} ||H_u||^4 + \lam_{dddd} ||H_d||^4 +  ||H_u||^2 ||H_d||^2 \left( \lam_{uudd}
 +   \cos^2\t(u,d) \left( \eta_{uddu} +  \xi_{udud} \cos2\phi(u,d) \right) \right) \nn \\
\eea
We know \cite{ma77} then the following conditions are necessary and sufficient for $V_{\mbox{\tiny Active}} \geq 0$
\bea
\lam_{uuuu} \geq 0 , \lam_{dddd} \geq 0 \label{lamuuuudddd>0} \\
\lam_{uudd} \geq -2 \sqrt{\lam_{uuuu} \lam_{dddd}} \label{lamuudd}\\
\lam_{uudd}+\eta_{uddu} \geq  |\xi_{udud}|  -2 \sqrt{\lam_{uuuu} \lam_{dddd}} \label{lamuudd+etauddu}
\eea

Now, for the coupled and inert parts of the potential we have, in the large field limit, a generic term of the form ($\lam_{a,b,c,d} H_a^\dagger H_b H_c^\dagger H_d$) where $a,b,c,d \in \{u,d,1,2,3,4\}$ with $\lam_{a,b,c,d} = \lam_{c,d,a,b}$ and $\lam_{a,b,c,d} \equiv \lam_{b,a,d,c}^*=\lam_{b,a,d,c}$.  We see that in order to insure  ($V_{\mbox{\tiny Coupled}} + V_{\mbox{\tiny Inert}}$) is positive, it suffices that any summation of the generic term with the terms corresponding to the permutations over the indices ($a,b,c,d$) is positive. More precisely, if we denote  the mapping $\sigma = \left( \begin{array}{cccc} 1 & 2 & 3 & 4 \\ a & b & c & d \end{array} \right) $ then the summand is of the form $\lam_{\sigma p(1), \sigma p(2), \sigma p(3), \sigma p(4)} H_{\sigma p(1)}^\dagger H_{\sigma p(2)} H_{\sigma p(3)}^\dagger H_{\sigma p(4)}$ where $p$ is a permutation in $S_4$ of $24$ elements. Noting that $\lam_{a,b,c,d} = \lam_{c,d,a,b}$, then $p$ can span $S_4/Z_2$ of $12$ elements, and using $\lam_{a,b,c,d} = \lam_{b,a,d,c}^*$ one can restrict the sum to six elements forming a set $\{p_k:k=1, \ldots, 6\}$ corresponding to ``independent'' couplings\footnote{For example, one can take \bea p_1=\left( \begin{array}{cccc} 1 & 2 & 3 & 4 \\ 1 & 2 & 3 & 4 \end{array} \right), p_2=\left( \begin{array}{cccc} 1 & 2 & 3 & 4 \\ 2 & 3 & 4 & 1 \end{array} \right), p_3=\left( \begin{array}{cccc} 1 & 2 & 3 & 4 \\ 2 & 1 & 3 & 4 \end{array} \right)  \nn \\
p_4=\left( \begin{array}{cccc} 1 & 2 & 3 & 4 \\ 1 & 3 & 4 & 2 \end{array} \right), p_5=\left( \begin{array}{cccc} 1 & 2 & 3 & 4 \\ 4 & 2 & 3 & 1 \end{array} \right), p_6=\left( \begin{array}{cccc} 1 & 2 & 3 & 4 \\ 2 & 3 & 1 & 4 \end{array} \right)
\eea.}, so that we impose
\bea &\sum_{k=1}^{k=6} 2 \mbox{Re} \lam_{\sigma p_k(1), \sigma p_k(2), \sigma p_k(3), \sigma p_k(4)} H_{\sigma p_k(1)}^\dagger H_{\sigma p_k(2)} H_{\sigma p_k(3)}^\dagger H_{\sigma p_k(4)} & \nn \\ &=& \nn \\
&2 \parallel H_a \parallel \parallel H_b \parallel \parallel H_c \parallel \parallel H_d \parallel \sum_{k=1}^{k=6} \lam_{\sigma p_k(1), \sigma p_k(2), \sigma p_k(3), \sigma p_k(4)}
\cos \t(\sigma p_k(1), \sigma p_k(2)) \cos \t(\sigma p_k(3), \sigma p_k(4)) \times& \nn \\ &   \cos \left[ \phi(\sigma p_k(1), \sigma p_k(2)) + \phi(\sigma p_k(3), \sigma p_k(4))\right] & \nn \\ & \geq 0 &
 \eea
 We see that if the independent coupling $\lam_{\sigma p_k(1), \sigma p_k(2), \sigma p_k(3), \sigma p_k(4)}$ is switched on, then one can always find a configuration where the phase exists and leads to a blowing downward. Two cases evading this instability happen either when all the indices $(a,b,c,d)$ are identical leading to $\t(a,a) =\phi(a,a)=0$, or when they consist of two different indices occurring each twice, such as $(a,a,b,b)$ or one of its permutations, where we get, noting that  $\t(a,b)=\t(b,a), \phi(a,b)= -\phi(b,a)$, the factor:
\bea 2 \parallel H_a \parallel^2 \parallel H_b\parallel^2 \left[ \lam_{a,a,b,b} + \cos^2 \t(a,b) \left( \lam_{a,b,b,a} +  \lam_{a,b,a,b} \cos 2\phi(a,b)\right)    \right] \eea

 Applying the above procedure on $V_{\mbox{\tiny Active}}$ (Eq. \ref{VCoupled}) we get the following sufficient conditions for its positivity in the large field limit.
  ($I \in \{u,d \}, \a \in \{1,2,3,4\}$)

\bea
\lam_{II \a \b} = \eta_{I \a \b I} = \xi_{I \a I \b} =0 , &\mbox{for}& \a < \b \nn \\
\lam_{II\a \a} \geq 0, \nn \\
\lam_{II\a \a}+\eta_{I \a \a I} \geq | \xi_{I \a I \a}| \nn \\
\Rightarrow \nn \\
V_{\mbox{\tiny Coupled}} \geq 0
\label{conditions_Coupled}
\eea
Note that here we did not get a relation relating, say, $\lam_{I,I,\a,\a}$ with $\lam_{I,I,I,I}$, because the term corresponding to the latter ($\lam_{I,I,I,I} ||H_{I}||^4$) has already been considered in deriving Eqs. (\ref{lamuudd}, \ref{lamuudd+etauddu}).

Applying similarly the procedure on $V_{\mbox{\tiny Inert}}$ (Eq. \ref{VInert}), we get the following sufficient conditions for its stability
\bea
\lam_{\a \a \a \a } \geq 0, \nn \\
\lam_{\a \a \a \b } = 0, &\mbox{for}& \a < \b , \nn \\
\lam_{\a \a \b \g } = \lam_{\a \b \g \a }= \lam_{\g \a \b \a}  = 0, &\mbox{for}& \a < \b < \g , \nn \\
\lam_{\a \b \g \d} = \lam_{\b \g \d \a} = \lam_{\b \a \g \d}= \lam_{\a \g \d \b} = \lam_{\d \b \g \a} = \lam_{\b \g \a \d}=0, &\mbox{for}& \a < \b < \g < \d , \nn \\
\lam_{\a\a \b\b} \geq 0 , &\mbox{for}& \a < \b , \nn \\
\lam_{\a\a \b\b} + \lam_{\a\b \b\a} \geq |2 \lam_{\a \b \a \b}|  , &\mbox{for}& \a < \b , \nn\\
\Rightarrow \nn \\
V_{\mbox{\tiny Inert}} \geq 0
\label{conditions_Inert}
 \eea
Note again that we did not seek to find a relation, as in Eqs. (\ref{lamuudd}, \ref{lamuudd+etauddu}), relating, say, $\lam_{\a\a\b\b}$ with $\lam_{\a\a\a\a}$, because if we impose the copositivity of the sum of terms corresponding to these two couplings ($\lam_{\a\a\a\a} \parallel H_a \parallel ^4 + \lam_{\b\b\b\b} \parallel H_\b \parallel^4 + \lam_{\a\a\b\b} \parallel H_\a \parallel^2 \parallel H_\b \parallel^2$, for $\cos \t(\a, \b) = 0$), then we can no longer use the term ($\lam_{\a\a\a\a} \parallel H_\a \parallel ^4$) to find a relation relating it to $\lam_{\a\a\g\g}$ with $\g \neq \b$.

 To summarize, if we take the conditions of Eqs. (\ref{lamuuuudddd>0}, \ref{lamuudd}, \ref{lamuudd+etauddu}, \ref{conditions_Coupled} and \ref{conditions_Inert}) then we guarantee that the potential is positive in the large field limit, and thus is bounded from below.

\subsection{Active Sector}
The mass spectrum in 2HDM has been given in some papers~\cite{ms}. We denote the two Higgses as follows.
\bea \label{ActiveHiggs} H_u = \bmat{c}  H_u^+\\ \phi_u+v_u+i \sigma_u\emat&,&H_d= \bmat{c}  H_d^+\\ \phi_d+v_d+i \sigma_d\emat \eea

The neutral CP-even fields ($\phi_u,\phi_d$) mix through a rotation matrix $Z^H$, characterized by an angle $\a$, to give mass eigenstates ($h_1,h_2$). The neutral CP-odd fields ($\sigma_u,\sigma_d$) equally mix through a rotation matrix $Z^A$ characterized by an angle $\b$ to give mass eigenfields ($A^0_1, A^0_2$) where $A^0_1$ is a massless Goldstone boson. Finally, the charged fields ($H^+_u,H^+_d$) will mix through a rotation matrix $Z^+$ characterized, up to a sign, by the same angle $\b$ to give mass eigenstates ($H^+_1, H^+_2,$) with massless Goldstone $H^+_1$. Thus we have five physical Higgses from the two doublets $H_u,H_d$:
 2 CP-even neutral ($h_1,h_2$)  with respective masses ($M_H=125\gev, M_e$), 1 CP-odd neutral ($A^0_2$) with mass $M_o$  and 2 charged Higgses ($H^+_2, H^-_2$) with mass $M_c$.

 We have seven free coupling parameters ($\mu_u^2, \mu_d^2, \lam_{uuuu}, \lam_{dddd}, \lam_{uudd}, \eta_{uddu}, \xi_{udud}$). We have also two vevs $v_u, v_d$ with $v_u^2+v_d^2=v_{\mbox{\tiny SM}}^2=246^2\gev^2$, determined by imposing the two tadpole conditions $\partial_{\phi_u} V_{\mbox{\tiny Active}}|_{\phi_u=v_u,\phi_d=v_d}=0, \partial_{\phi_d} V_{\mbox{\tiny Active}}|_{\phi_u=v_u,\phi_d=v_d}=0$. We define $\tan \b = \frac{v_d}{v_u}$.

 We shall tradeoff our free parameters with the following physical parameters: ($M_H=125,M_{e}, M_{o}, M_{c}, \a, v=v_{\mbox{\tiny SM}}, \tan \b$).

 In Appendix (\ref{Mass Matrices for Scalars}), we state all the related formulae in the enumerations (\ref{Mass matrix for Higgs}, \ref{Mass matrix for Pseudo Higgs}, \ref{Mass matrix for Charged Higgs}).

\subsection{Inert Sector}
We denote the additional Higgs doublets ($\a=1,2,3,4$) by:
\bea H_\a &=& \bmat{c}  H_\a ^+\\ \phi_\a +i \sigma_\a \emat\label{InertHiggs}\eea

Again, the neutral CP-even (odd) fields $\phi_\a$ ($\sigma_\a$) mix amidst themselves with rotation matrix $ZH^i$ ($ZA^i$). Equally, the charged fields $H_\a^+$ mix together with rotation matrix $ZP^i$

In Appendix (\ref{Mass Matrices for Scalars}), we state all the related formulae in the enumerations (\ref{Mass matrix for Inert Higgs}, \ref{Mass matrix for Inert Pseudo Higgs}, \ref{Mass matrix for Inert Charged Higgs}). The scalar particle content appears in Table form in Appendix (\ref{Particle Contents}).

\section{``Symmetric" Benchmark Point}
In some regions of the 2HDM
 parameter space
there is a decoupling limit in which the masses of the CP-odd and
charged fields are quite larger
 than those of two CP-even. Because our purpose is to see whether or not the (LAH) in our model
 can be a component of the DM, we limit ourselves for simplicity to this decoupling
 limit, as we shall now explain. Also, for the ``Inert'' section we opt for a simplifying choice which amounts to decouple one ``light'' doublet (say the $1$st) from the other ``heavy'' doublets, as will become clearer shortly. The benchmark point we end up resembles thus a special (2+1) model which is worthy to be investigated on its own merit.

\subsection{Active Sector Benchmark Point }
We fix our parameter space at an experimentally accepted point known as the  ``Alignment Limit" satisfying ($\sin (\b - \a ) = 1$) \cite{das}.

Furthermore, we go for a ``symmetric'' alignment limit in the sense we take:
\bea \label{symmetric alignment limit} \a = -\b = -\pi/4 &\Rightarrow& \tan \b = 1\eea
We note that the choice $\a=0$ is rejected by the current data \cite{das}.
Finally, we decouple the CP-odd and charged Higgses from the CP-even Higgs, and take a ``democtratic'' limit, acceptable according to \cite{das}, reducing the parameters number and mimicking the IDM:
\bea \label{symmetric choice Active} M_{h_1} = M_{h_2} \equiv M_H = 125 \gev &,& M_o, M_c \gg 125 \gev\eea

In fact, there is a limit in the 2HDM parameter space corresponding to degenerate CP-even Higgs fields decoupled from heavy CP-odd and charged Higgses, which was, together with other decoupling degenerate points, thoroughly studied in \cite{Haber_2013} and contrasted to the LHC data. To see this, it suffices to ``fine tune'' and
take the following constraints on the lagrangian parameters:
\bea \label{constraints}
\left|\frac{\lam_{uuuu}}{\xi_{udud}}\right| = {\cal{O}}(\eps) &,& \left|\frac{\lam_{uuuu}}{\xi_{udud}+\eta_{uddu}}\right|= {\cal{O}}(\eps) , \nonumber \\
\left|\frac{\xi_{udud}+\eta_{uddu}+\lam_{uudd}}{\lam_{uuuu}}\right|=\left|{\cal O}(\frac{1}{\eps})+\frac{\lam_{uudd}}{\lam_{uuuu}} \right|= {\cal{O}}(\eps)  &,& \left|\frac{\lam_{uuuu} - \lam_{dddd}}{\lam_{uuuu}}\right|=\left|1-\frac{\lam_{dddd}}{\lam_{uuuu}}\right|={\cal{O}}(\eps^2)
\eea
with $\eps \ll 1$, then with the choice $\tan{\b}=1$ we see from Eq. (\ref{m2evenspectrum}, \ref{m2oddspectrum}, \ref{m2chargedspectrum}) that $M_0, M_c \gg M_{h_1} \approx M_{h_2}$ and that $|\a|=|\b|=\pi/4$.\footnote{We shall state our results to zeroth order in $\eps$}. Note here that it is the combination $(\xi_{udud}+\eta_{uddu}+\lam_{uudd})$ which enters as a perturbative parameter in the unitarity constraints. Note also that $M_o, M_c$, albeit being large, should not be too heavy in a way to affect the gauge couplings unification.

\subsection{Inert Sector Benchmark Point}
We set now our benchmark point in the ``inert'' sector with many simplifying assumptions. As we shall see, these ``strong'' assumptions lead to singling out the LAH easily, and to decouple it from the rest of the spectrum. Our objective is to reduce the number of parameters, and see
whether or not the recovered LAH  can account for DM.

First, we assume all the off-diagonal elements of the mass matrices for CP-even (odd) and charged Higgs fields (Eqs. \ref{mphiphi}, \ref{sigmasigma} and \ref{h+h+}) are zero. Thus, we have:
\bea
m^2_{F} &=& \mbox{diag}\left( M^{F}_{11}, M^{F}_{22}, M^{F}_{33},M^{F}_{44}\right) , \label{mass matrix generic inert}.
\eea where $F\in\{h^i, A^{i0}, H^{i+}\}$.
Second, we assume that only the first inert doublet is coupled to the active sector, i.e.
\bea
\label{firstOnlyWithActive}
\mbox{For} I=u,d, \a=2,3,4 &,& \lam_{II\a\a}=\eta_{I\a\a I}=\xi_{I\a I\a}=0 .
\eea
Looking again at Eqs (\ref{mphiphi}, \ref{sigmasigma} and \ref{h+h+}) we have then
\bea M^F_{\a\a} &=& \m_{\a\a}, \a=2,3,4\label{malphaalpha}\eea
and
\bea
M^2_{e1}\equiv M^{h^i}_{11}&=&\m_{11}+ (\lam_{uu11}+\eta_{u11u}+\xi_{u1u1}) v_u^2 + (\lam_{dd11}+\eta_{d11d}+\xi_{d1d1} )v_d^2
\label{Mh11},\\
M^2_{o1} \equiv M^{A^{i0}}_{11}&=&\m_{11}+ (\lam_{uu11}+\eta_{u11u}-\xi_{u1u1}) v_u^2 + (\lam_{dd11}+\eta_{d11d}-\xi_{d1d1} )v_d^2
\label{MA11},\\
M^2_{c1} \equiv M^{H^{i+}}_{11}&=&\m_{11}+ (\lam_{uu11}) v_u^2 + (\lam_{dd11})v_d^2
\label{MH+11},
\eea
 Third, we assume that $\phi_1$ decouples from $\phi_\a$ for $\a=2,3,4$, by imposing
 \bea \m_{\a\a}/\m_{11}&\gg& 1,\a=2,3,4 \eea
As to $M^F_{11}$ (Eqs. \ref{Mh11}, \ref{MA11} and \ref{MH+11}), we see that we have seven parameters. Fourth, we impose an u-d symmetric condition:
\bea \lam_{uu11} = \lam_{dd11},\eta_{u11u}=\eta_{d11d} &\mbox{and}& \xi_{u1u1}=\xi_{d1d1} \label{udsymmetric}\eea
Thus we end up with four parameters which determine the elements $M^F_{11}$. If we denote now
\bea \lam = \lam_1 \equiv \lam_{uu11}+\eta_{u11u}+\xi_{u1u1} &=& \lam_2 \equiv \lam_{dd11}+\eta_{d11d}+\xi_{d1d1}\label{lambda},\eea
 then we can tradeoff the four free parameters by the masses $M_{e1}, M_{o1}, M_{c1}$ and the constant $\lam$ expressing the coupling strength between the DM and the active Higgs field.
 Finally, and as a fifth assumption similar to our assumption in the 2HDM, we decouple CP-even inert Higgs from the CP-odd and charged inert Higgses by assuming:
 \bea M_{o1}, M_{c1} &\gg& M_{e1}\label{decoupling}\eea
 Looking at Eqs. (\ref{Mh11}, \ref{MA11} and \ref{MH+11}), we see that this assumption means restricting ourselves to a region in the parameter space where $\left|\lam / (\lam + \eta + \xi)\right| \gg 1, \left|(\lam + \eta - \xi) / (\lam + \eta + \xi)\right| \gg 1$ but $\left|\lam + \eta + \xi\right|$ is finite. Note again that it is the combination $(\lam + \eta + \xi)$ which is involved in the vertex between physical states and which should be perturbative in order to respect the constraints of unitarity.

 We call the point which satisfies the above five assumptions in the inert sector by the ``Symmetric Dark'' benchmark, as it corresponds to symmetric couplings of $H_u$ and $H_d$ to the remaining light inert doublet containing our DM candidate. At the ``symmetric dark'' benchmark, the free parameters of the dark sector are determined only by two free parameters $M_{e1}\equiv M_{DM}, \lam$, putting aside the self-coupling $\lam_{1111}$. Now the LAH, referred to by $\psi$ is nothing but the lightest neutral component of the inert doublet $h^i_1$. The simplicity of the ``Symmetric Dark'' benchmark point and the reduced number of its free parameters to be scanned is what pushed us to consider it for the sake of demonstrating the model viability.

\subsection{Total Benchmark Point}

Our ``symmetric choice'' benchmark is the the ``symmetric alignment limit" of the active sector ($v_u=v_d$) PLUS the ``symmetric Dark" benchmark point. Thus we have
\bea M^2_{DM}=M^2_{e1}=M^2_{\psi}&=&\m_{11}+\lam v^2\eea

The last column of the Table in Appendix (\ref{Particle Contents}) shows the mass decoupling at the ``Symmetric choice" benchmark \footnote{ Upon request, we can provide the Feynman rules for the vertices at the chosen benchmark point.}.

 Although our chosen point in (2+4) scenario can not be distinguished, as long as dark matter is concerned, from the (2+1) scenario, nonetheless, we need six Higgs doublets for unification purposes.

\section{Phenomenology}

As said earlier, we restricted the phenomenolgical analysis to our ``symmetric choice" benchmark point, as the stress in this work is on model building and its viability for accommodating DM. However, this benchmark is important in itself, as it represents a special case of the (2+1) model when $M_{h_1}=M_{h_2}$. However, it is not equivalent to the IDM, as the active Higgses can be differentiated by their decay products and can not be identified both to the LHC Higgs, in that $h_1$ can not decay directly to gauge bosons in contrast to $h_2$.

We used for the numerics the Mathematica packages ($Sarah+FeynArts+FormCalc$), where we computed explicitly all the relevant quantities, with no need to go to some ``blackbox" packages, such as micrOMEGAs. We made use, however, of the latter package
in order to check consistency with mathematica results concerning the different adopted constraints.

\subsection{Relic Density}

An appealing aspect of WIMPs as DM candidate is that it could be
produced naturally with an abundance of approximately the correct
order of magnitude \cite{khalil}. By using the current observational data on DM
abundance, further constraints on the parameter space can be
obtained.

If we assume that the DM is produced in a thermal process, and the
production process ceases before annihilation, then
its abundance evolves as
\bea \frac{dn_\psi}{dt}  =
-3Hn_\psi - \langle \sigma_{\rm tot} v_{\rm rel } \rangle
                    (n_\psi^2 - (n_\psi^{\rm eq})^2).
\label{b} \eea where $H$ is the Hubble parameter, $n_\psi^{\rm
eq}$ is the equilibrium density of the particle $\psi$, and
$\langle \sigma_{\rm tot} v_{\rm rel } \rangle$ is the thermal
average of the annihilation cross section. The DM abundance
freezes as annihilation rate drops below the Hubble rate, and if we denote $Y_{\infty}$ as the asymptotic
ratio of DM number density to entropy density, then upon expanding  the
cross section to first order in powers of $x^{-1}\equiv
\frac{T}{m_\psi}$, so that to write $<\sig_{\rm tot} v_{\rm rel}>
= \sig_0 (1+b\: x^{-1})$, then $Y_{\infty}$ is given by \cite{kolbturner} \bea
Y_\infty &=& \frac{n_{\psi 0}}{s_0}= \frac{3.79
x_f}{(\frac{g_{\star S}}{\sqrt{g_\star}})m_{PL}m_\psi \sig_0
(1+\frac{b}{2 x_f})} \eea with
\begin{dmath} \label{x_f}
x_f = \log\left[ 0.038 \frac{g_{\psi}}{\sqrt{g_\star}} m_{pl} m_\psi \sigma_0 \right] - \frac{1}{2} \log\left[ \log\left[ 0.038 \frac{g_{\psi}}{\sqrt{g_\star}} m_{pl} m_\psi \sigma_0 \right] \right] + \log\left[ 1+ \frac{b}{\log\left[ 0.038 \frac{g_{\psi}}{\sqrt{g_\star}} m_{pl} m_\psi \sigma_0 \right]} \right]
\end{dmath}
 where $g_\psi = 1$ for the scalar boson $\psi$. As to $g_\star$, it counts the total number of effectively mass degrees of freedom (for species with mass far less than the temperature), whereas $g_{S\star}$ expresses a proportionality between the entropy density and $T^3$, and for all particle species, except neutrino, can be replaced by $g_\star$ for most of the history of universe. Thus, we get finally
 \cite{kolbturner}:
\bea \label{Omegah2}
\Om_\psi \equiv \frac{n_\psi m_\psi}{\rho_c} &\Longrightarrow&
\Omega h^2 = \frac{1.07 \times 10^{9} x_f \gev}{\sqrt{g_\star} m_{pl} \sigma_0 \left(1+\frac{b}{2 x_f} \right)}
\eea

With our model we may calculate explicitly the annihilation cross section of
$\psi$. We shall include only tree diagrams, and consider only
annihilations involving 2-body final states, as final states
involving three bodies or more are suppressed by the phase space
factor. The Feynman diagrams responsible for such interaction can
be classified into two categories: the first involves 4-point
vertices leading to gauge bosons or 2HDM Higgses,, and the second has an s-channel for an
intermediate Higgs ($h_1$ or $h_2$) or a t(u)-channel with an intermediate $\psi$. At the tree level, the annihilation can not
proceed via intermediate gauge bosons because of the absence of
$\psi\psi W(Z)$ vertices. For completeness, we state in Appendix \ref{annihilation} all the annihilation amplitudes where we used the FaynArts, FormCalc packages in order to compute the polarization sum for bosons, and the helicity/color sum for fermions.

We have used the method of \cite{olive94,olive88,sredinski} for taking the thermal average. Precisely, we use the following formulae for the thermal average of annihilation of initial two bodies with masses $m_1,m_2$:
\be \label{1}
\langle \sigma v_{rel} \rangle = \frac{1}{m_1 m_2} \left( 1-\frac{3(m_1+m_2)T}{2m_1 m_2} \right) w(s)
\ee
where:
\bea \label{2}
w(s) = \frac{1}{32 \pi} \frac{p_f(s)}{s^{1/2}} \int_{-1}^{+1} d\cos \theta_{CM} |\mathcal{M}|^2 \\
s = (m_1 + m_2)^2 + 3 (m_1 + m_2) T
\eea
where $p_f(s)$ is the magnitude of the three-momentum of one of the final particles in the center-of-mass frame.

We determine the freezing temperature $T_f$ by iteration. For a given $T$, we compute the thermal average $\langle \sigma v_{rel}\rangle$ using Eqs. (\ref{1}, \ref{2}), which we expand in powers of $T$ to determine the coefficients $\sigma_0,b$. We compute $g_\star(T)$ using the formula \cite{kolbturner}:
 \bea \label{g_*} g_\star&=& \Sigma_{i=\mbox{boson}}g_i+\frac{7}{8}\Sigma_{i=\mbox{fermion}}g_i\eea
 where $g_i$ expresses the number of internal degrees of freedom of the species $i$ with mass $m_i$ less than the temperature $T$, then we compute the ``new" temperature using Eq. (\ref{x_f}). We iterate till we get convergence.

We note also that in order to determine the coefficients of the expansion $\langle \sigma v_{rel}\rangle (T) = \Sigma_n (\mbox{coef.})_nT^n$, the function $\langle \sigma v_{rel}\rangle (T)$ is not a continuous function. Thus, we can not by simple differentiation determine the linear term. What we did is to find the best fit of a linear term by using the method of Least Square Error in the range $T\leq T_f$ since the thermal average formulae are valid only for this range \cite{sredinski}.

We draw the attention that in our explicit analysis, one should not expect a complete agreement with the more sophisticated packages like micrOMEGAs. In fact, we know \cite{gondolo} that expanding the thermal average $\langle \sigma v_{rel}\rangle_T = \Sigma_n (\mbox{coef.})_nT^n$ is not valid at resonances and/or at thresholds, and one should take account of many effects including, say, the Sommerfeld enhancement. For the resonance in our case, we have the kinematically allowed limit $M_{\psi}=\frac{1}{2} M_H$ and our resonance is limited to Higgs, since DM can only annihilate via Higgs. Equally, one should amend the formula around the thresholds $s=4M_f^2$ for a final product of mass $M_f$. However we checked partially when comparing to micrOMEGAs that the effects due to these changes are small and do not change much the phenomenology. Actually, the micrOMEGAs package uses corrected formulae, and although our initial analysis may give different results around resonances and thresholds, but we checked locally around many chosen points at this critical region that discrepancies are numerically small. Since, in our scan, we took $M_{\psi} > 2 M_q$ for all light quarks, so the first threshold effect should appear at the masses of the $Z, W^+$ gauge bosons.

Moreover, in our analysis, there is no role for co-annihilation, as there is no particle in the dark sector with mass close to $M_{DM}$ in our chosen ``symmetric" benchmark point.

Phenomenologically,  we did not equate the DM relic density to a given value and solved, at fixed $M_{DM}$ ($\lam$), for $\lam$ ($M_{DM}$). Rather, we fixed $M_{DM}$ ($\lam$), and plotted the DM relic density as a function of $\lam$ ($M_{DM}$). We can now determine and visualize the accepted region of $\lam$ ($M_{DM}$)  for the taken fixed value of $M_{DM}$ ($\lam$) by imposing a band condition on the DM relic density. In fact we have scanned a 2-dim parameter space of $\lam$ and $M_{DM}$, spanning 300000 points on a simple core i5 intel processor, and for each parameter space point we imposed the band condition.

\begin{figure}[hbtp]
\centering
\includegraphics{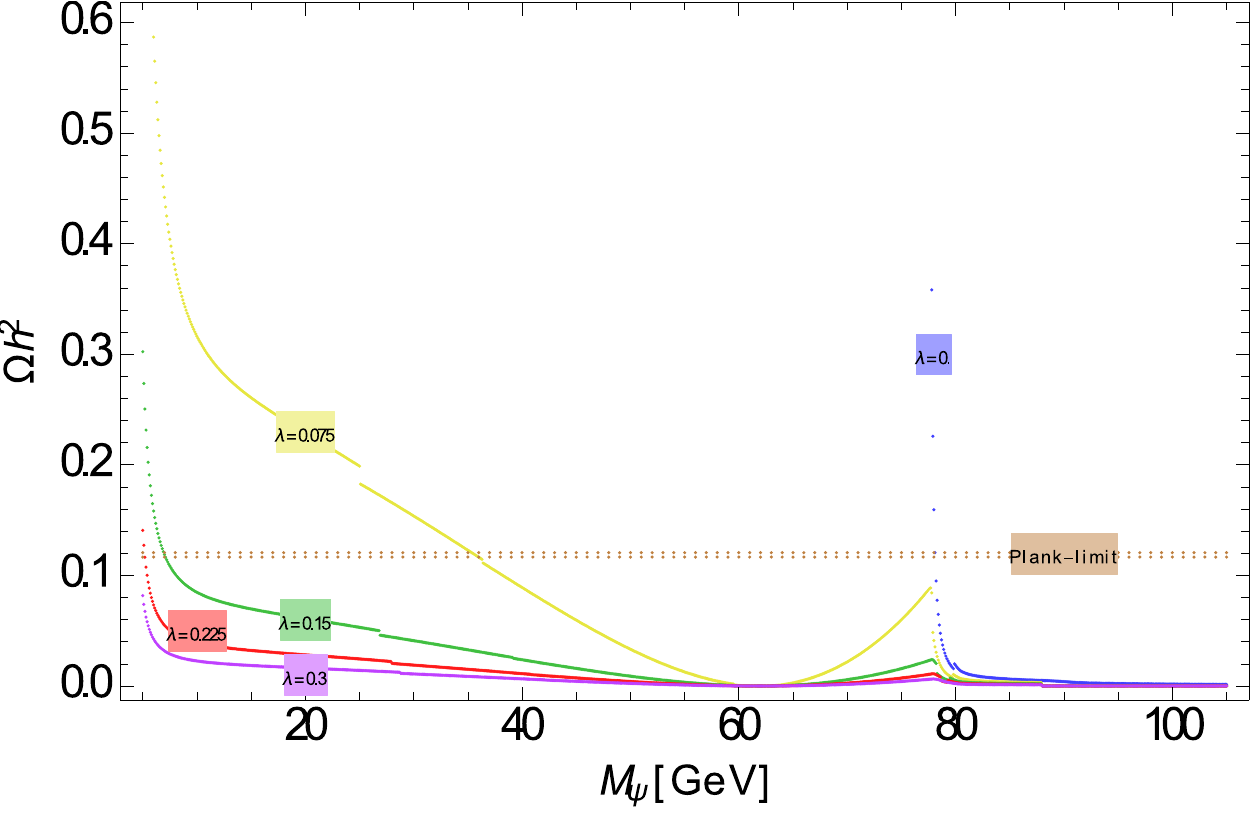}
\caption{{\footnotesize Relic Density versus $M_\psi$ for 5 fixed values of $\lam$. ``Acceptable'' Planck limits are also shown.}}
\label{fig1}
\end{figure}

\begin{figure}[hbtp]
\centering
\includegraphics{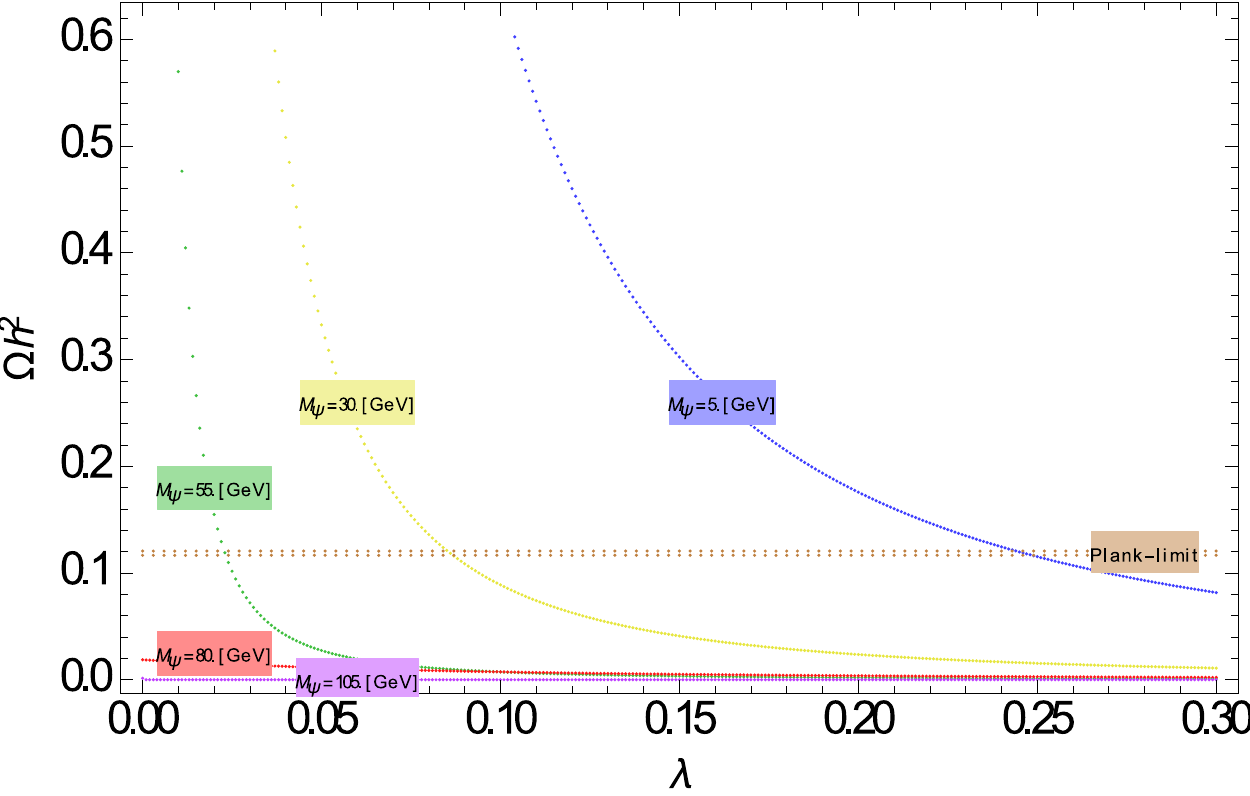}
\caption{{\footnotesize Relic Density versus $\lam$ for 5 fixed values of $M_\psi$. ``Acceptable'' Planck limits are also shown. }}
\label{fig2}
\end{figure}

In Fig. \ref{fig1}(\ref{fig2}), $\Omega$ is plotted versus $M_\psi$ ($\lam$) for 5 fixed values of $\lam$ ($M_\psi$).

\subsection{Direct Detection}

The connection between $\lam$ and $m_\psi$ derived from
the abundance constraint is very predictive. It has consequences
for current DM searches. The sensitivity of DM detectors to $\psi$
particles is controlled by their elastic scattering cross section
with visible matter, and in particular with nuclei.
In fact, this
cross section is the relevant quantity to be measured in the
current experiments designed to measure the recoil signal of DM
collisions within detectors \cite{Lux,xenon}. More specifically, we shall state the Direct Detection (DD) constraints at the level of scattering off
nucleons. Actually, there are well known formulae starting from the nucleon's ones in order to compute the DD cross section between DM and nuclei \cite{ragu}.

Also, the
elastic DM-nucleon scattering controls the abundance of DM
particles trapped at the terrestrial or solar core, and whose
presence is detected indirectly through the flux of energetic
neutrinos which is produced by subsequent DM annihilations. However, in this work, we shall not pursue the viability of the model  with respect to indirect detection.

The Feynman diagram describing elastic $\psi$-particle collisions
with nucleons corresponds to a
T-channel via an intermediate 2HDM Higgs $h_1, h_2$. We assume
loop diagrams exchanging gauge bosons to be neglected because of the presence of
more than one propagator, contrary to tree-level diagrams.

We use the non-relativistic approximation for the incoming
particles, and thus apply low energy theorems to have an effective
Higgs-Nucleon coupling \cite{Higgs}\bea \label{chiral} \eta_{\mbox{N}}\equiv g_{HNN} &\approx & \frac{g
M_N}{2 M_W}\frac{2 N_H}{3 q} \eea where $M_N \approx 939 MeV$ is
the nucleon mass, $N_H$ is the number of heavy quarks whose masses
exceed the Higgs mass and $q = 11-\frac{2}{3}n_L$ with $n_L=6-N_H$
the number of light quarks. We get in Appendix (\ref{DDamplitudes}) a scattering amplitude given by Eq. (\ref{M2DD}) and a thermal average for the
elastic $\psi - N$ scattering given by Eq. (\ref{sigmavDD}).

The LUX experiment \cite{Lux} puts an upper bound $2.2 \times 10^{-46}$ $\mbox{cm}^2$ on $\langle \sigma_{\mbox{el}} v\rangle$, whereas the Xenon100 experiment \cite{xenon} puts a larger upper bound ($1.1 \times 10^{-45}$ $\mbox{cm}^2$).

\begin{figure}[hbtp]
\centering
\includegraphics{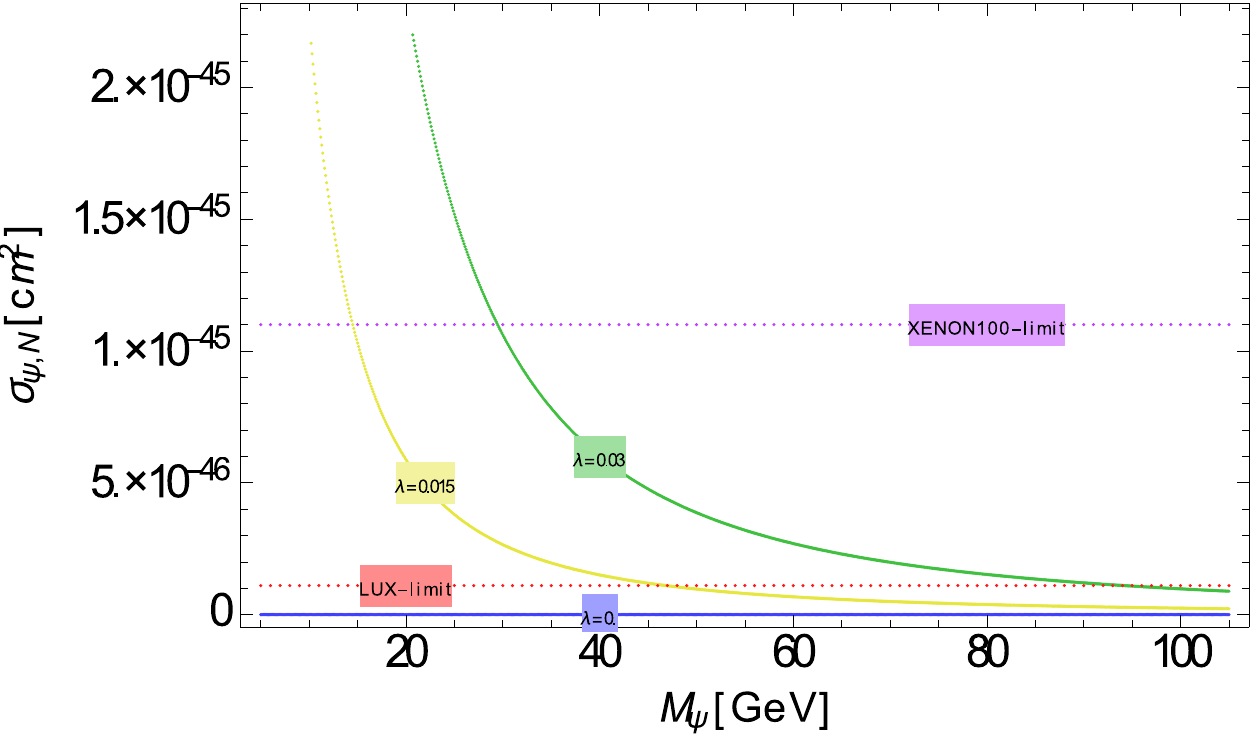}
\caption{{\footnotesize Direct Detection: Nucleon-$\psi$ scattering amplitude versus $M_\psi$ for 3 fixed values of $\lam$. LUX and Xenon100 limits, above which lie excluded regions, are also shown.}}
\label{fig3}
\end{figure}

\begin{figure}[hbtp]
\centering
\includegraphics{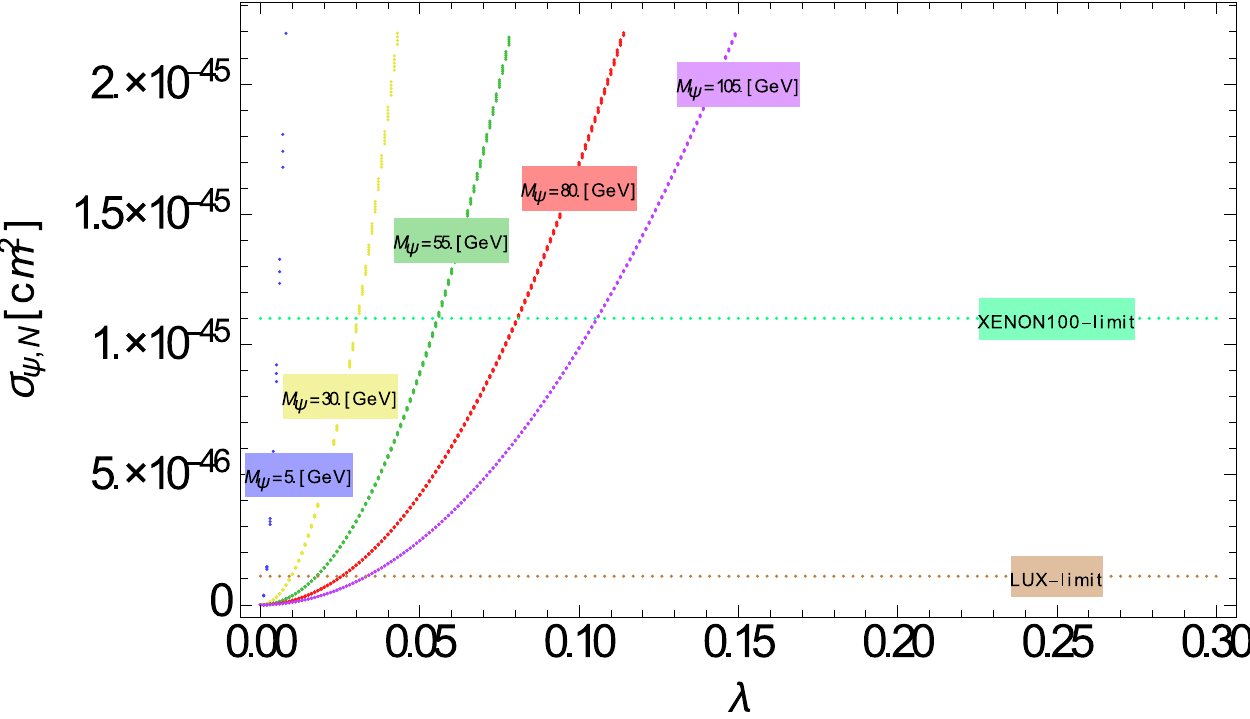}
\caption{{\footnotesize Direct Detection: Nucleon-$\psi$ scattering amplitude versus $\lam$ for 5 fixed values of $M_\psi$. LUX and Xenon100 limits, above which lie excluded regions, are also shown.}}
\label{fig4}
\end{figure}

In Fig. \ref{fig3}, we plot $\sigma_{\mbox{el}}$ versus $M_{\psi}$ for a fixed value of $\lam$, and we show the LUX limit on the figure showing that for various decreasing values of $\lam$ considerable augmenting ranges of $M_\psi$ pass the DD constraint. In contrast, in Fig. \ref{fig4} we plot $\sigma_{\mbox{el}}$ versus $\lam$ for five fixed value of $M_{\psi}$, and we see
that the acceptable range of $\lam$ shrinks as $M_\psi$ decreases.

When we scan the 2-dim parameter space of ($M_\psi, \lam$)
we see that LUX experiment allows just a small region around resonance (DM is light around 62 GeV), whereas for Xenon100 there is a much larger portion which is not excluded. \footnote{One should note however that the approximation of Eq. (\ref{chiral}) is a crude one, and, in a more refined analysis, tougher exclusion curves, resulting from the experimental data, are expected. However, the study of \cite{Ghosh} argued that the interplay of annihilation, co-annihilation and semi-annihilation processes relax the DD constraints in various non-minimal scalar Higgs portal DM models.}

\subsection{Invisible Higgs Decay}
As said earlier, one can distinguish $h_1$ from $h_2$ by their decay products. \footnote{$h_u, h_d$ become $h_1, h_2$ mass Eigenstates. In our benchmark $M_{h_1} = M_{h_2} = M_{higgs}$. However,  the vertex $(h_2, \psi, \psi) \neq 0$ whereas the vertex $(h_1, \psi, \psi) = 0$.}. The Higgs $h_2$ can decay via 4 channels $\psi \psi$, $f  \bar{f}$, $W^+ W^-$ and $Z,Z$ whereas the Higgs $h_1$ decays only to $f \bar{f}$. We identify the LHC Higgs by $h_2$ and take the Invisible Higgs Decay (IHD) constraint to be:
\bea |\frac{h_2 \rightarrow DM+DM}{h_2 \rightarrow X}|^2<0.2 \label{IHDconstraint}\eea
For completeness, we list in Appendix \ref{IHDappendix} all the IHD amplitudes.

\begin{figure}[hbtp]
\centering
\includegraphics{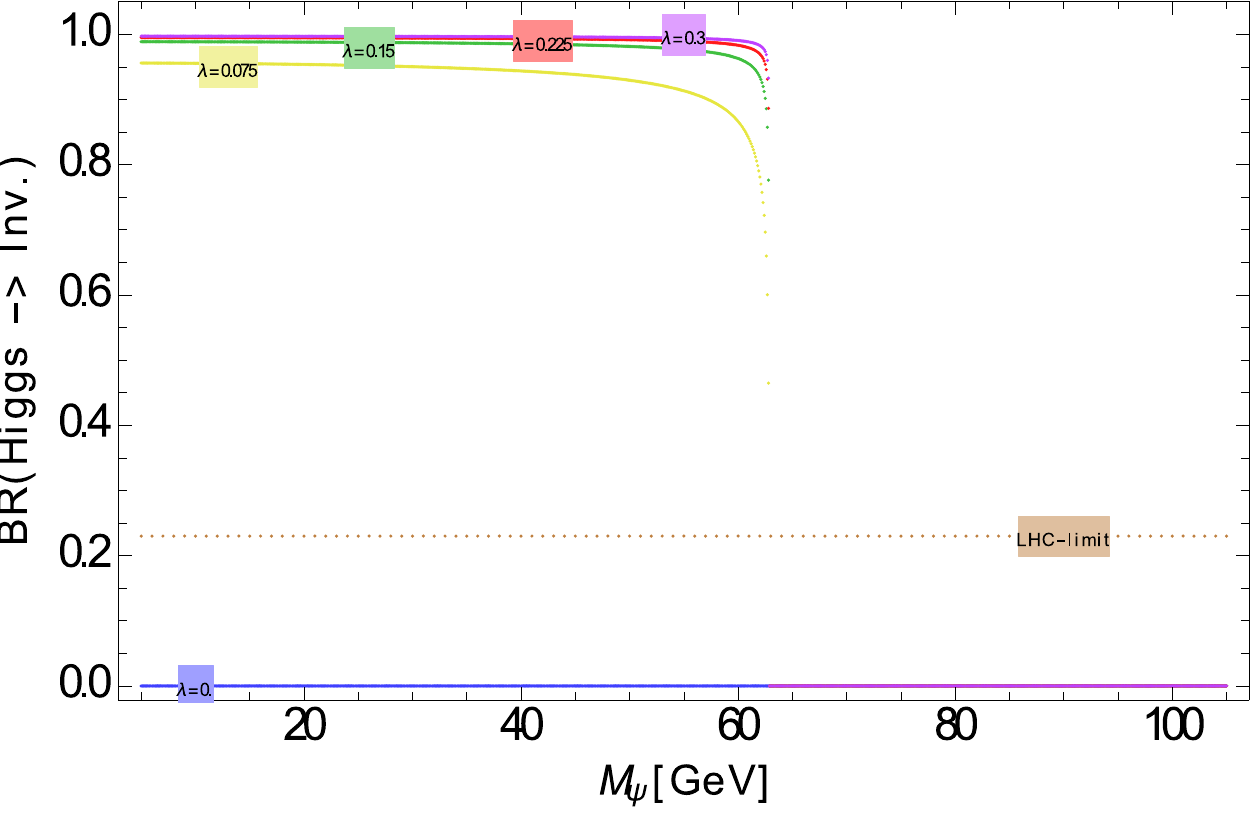}
\caption{{\footnotesize Branching Ratio of Invisible Higgs Decay versus $M_\psi$ for 5 fixed values of $\lam$. The LHC limit, above which is an excluded region, is also shown. }}
\label{fig5}
\end{figure}

\begin{figure}[hbtp]
\centering
\includegraphics{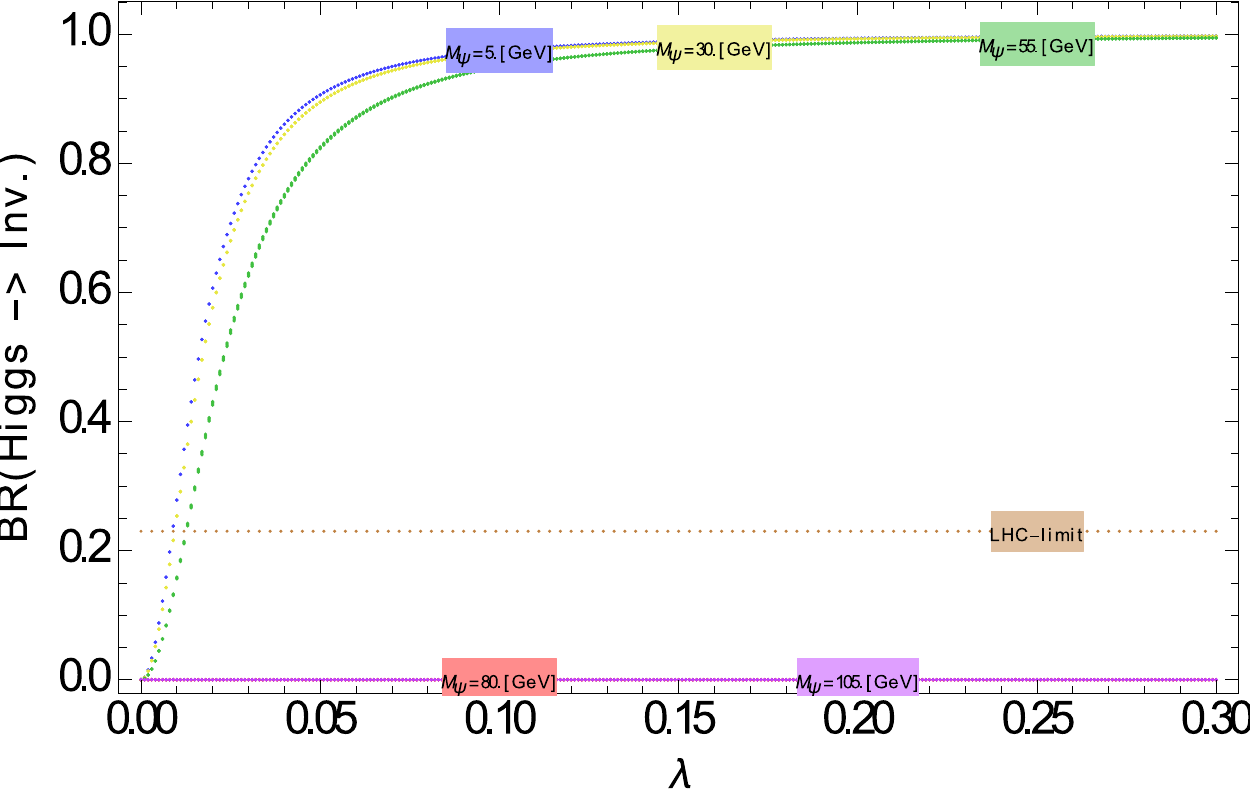}
\caption{{\footnotesize Branching Ratio of Invisible Higgs Decay versus $\lam$ for 5 fixed values of $M_\psi$. The LHC limit, above which is an excluded region, is also shown.}}
\label{fig6}
\end{figure}

In Figs. (\ref{fig5}, \ref{fig6}), we show respectively the IHD ratio vs ($m_\psi, \lam$) showing acceptable points with respect to the IHD constraint.

\subsection{Mean Free Path}
We estimate here the mean free path $l_{\psi}$ of the DM, which should be larger than $1~\mbox{kpc}$, otherwise
DM particles would behave as a collisional
gas altering substantially the halo structure and evolution. We
have: \bea l_{\psi} = {1 \over \sigma_{\psi \psi} n_\psi} =
{m_\psi \over \sigma_{\psi\psi} \rho^{h}_\psi} ~~, \eea where
$n_\psi$ and $\rho^{h}_\psi$ are the number and mass density in
the halo of the $\psi$ particle, respectively, and we use
$\rho^{h}_\psi \approx 0.4$ GeV/cm$^3$, corresponding to the halo
density.

The amplitude of the elastic $\psi-\psi$
scattering is stated in Appendix \ref{self elastic scattering amplitudes} (Eq. (\ref{M2SES}))\footnote{Loop diagrams involving gauge bosons are dropped numerically. This comes because a such
diagram would be accounted for by a renormalized self-coupling
$\lam_S^R$ and a logarithmic term. The latter, on dimensional
grounds, would be proportional to $g^4 \ln\frac{s}{m_W^2}$ to be
compared with the intermediate Higgs s-channel diagram which is
proportional to $\lam^2\frac{m_W^2}{m_H^2}$, with the
total energy-squared $s$ proportional to $m_\psi ^2$. One can see
now that, since $m_\psi^2, m_W^2, m_H^2$ fall all in the weak scale,
there are numerical choices making the logarithmic term far less
than the tree-level corresponding term.}. We note that it is only here where the ``self''-coupling ($\lam_S \equiv  \lam_{1111}$) interferes.
 Using $t=-2 (1-cos \theta)
(\frac{s}{4}-m_\psi^2), u=-2 (1+cos \theta)
(\frac{s}{4}-m_\psi^2)$ with $\theta$
 is the scattering angle in the center of mass frame, we get Eq. (\ref{sigmavSES}), in Appendix \ref{self elastic scattering amplitudes}, expressing the thermal average of the self-elastic scattering cross section $\langle \sigma_{\mbox{el}}\rangle$.

We can plug in values for $m_\psi,\lam$ corresponding to points which have passed the constraints of relic density, DD and IHD,
and compute $\sig_{\rm el}$ and $l_{\psi}$ for some choices of the
DM self-coupling $\lam_S$, which is completely unconstrained by
the relic abundance condition.

For a ``generic" choice of parameters, with a `perturbative' value
for $\lambda_S$, we typically get a mean free path of the order
of $10^{7} \mbox{Mpc}$ or larger, similar to other CDM candidates. It is well
known that the CDM model has the so called ``missing satellite"
problem \cite{problem}, though its solution may either originate from the distinctive property of
the dark mater particles (e.g. warm dark matter or strongly
self-interacting dark matter), or be due to astrophysical reasons
such as negative feedback of star formation in the small subhalos.
We note that while the generic 6HDM model is in this respect
similar to other CDM models, the model in special circumstances
may provide some mechanism to help solve this problem, e.g. the
mean free path may be shortened (look at Eq. \ref{sigmavSES}) with a very large self coupling
$\lambda_S$, or pushing down the mass of $\psi$ to the $\mbox{Mev}$ range.
Also, we have considered here $\psi$ as the dominant contribution
to DM, but there may be scenarios where it is only part of the DM, which
could open ways to solve this galactic-scale problem.

\section{Conclusions}
\begin{figure}[hbtp]
\centering
\includegraphics[scale=0.9]{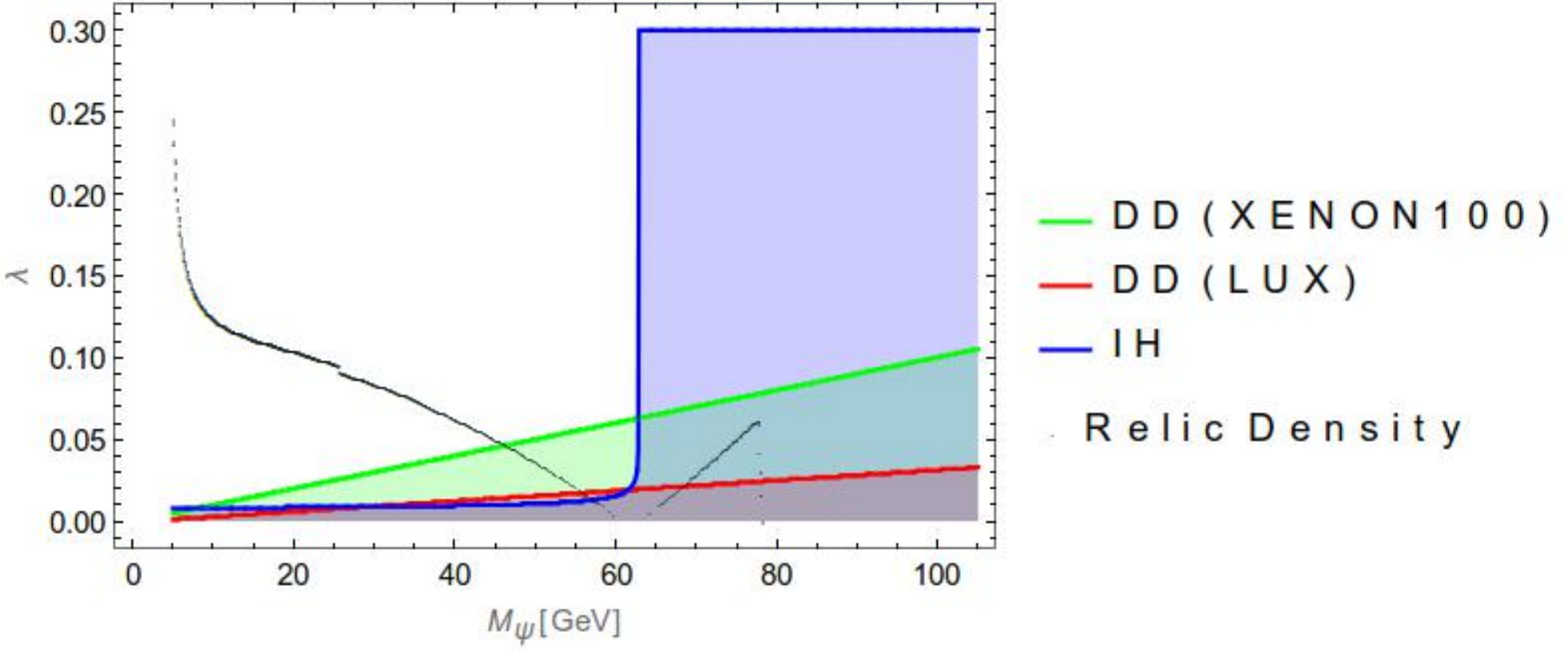}
\caption{{\footnotesize 6HDM at the ``symmetric" choice for benchmark point. }}
\label{fig7}
\end{figure}

 Fig. \ref{fig7} represents the phenomenological analysis of the (2+4) model at the ``symmetric choice'' benchmark point. The two axes represent our 2-dim parameter space spanned by ($m_\psi,\lam$), where $\lam$ is restricted to be ``perturbative" ($\lam \leq 1$). The ``black'' band represent the points which satisfy the RD constraint \cite{particledata}:
  \bea \label{RelicConstraint} \Omega h^2 \in  0.1186  \pm 0.0020 \eea

Although we scanned $m_\psi$ in the range $[5 \mbox{ GeV}, 1 \mbox{ TeV}]$, we could only find points satisfying the RD constraint in the range $m_\psi < 100 \mbox{ GeV}$, to which we restricted the horizontal axis in Fig. \ref{fig7}.

  The ``red'' line represents the DD LUX limit, above which is to be excluded. We also state the DD Xenon100 limit (``green'' line) which excludes less regions than the LUX limit. The ``blue" line represents the IHD constraint keeping the region underneath. We see that there is a viable region in parameter space around the resonance ($m_H/2$) satisfying all the constraints.

Looking at the black relic density band, we see two decreases around the resonance ($m_H/2$) and the threshold ($m_Z\simeq m_W$). This is to be expected as,
in general terms, one expects that the annihilation cross section would increase around the resonance. In order to get non-negligible DM relic density, we need thus to decrease the coupling $\lam$ around the resonance. Similarly, and because of the channel $DM+DM\rightarrow WW$, we expect a decrease in $\lam$ when $M_{\mbox{DM}}$ is around $M_W$.
In our ``symmetric" benchmark we have just one resonance $M_{H_u}=M_{H_d}=M_H\Rightarrow M^{\mbox{resonance}}_{DM}=\frac{1}{2} M_H$, whereas we expect for other choices where $M_{h_1} \neq M_{h_2}$ to get two resonances.

We also noted a ``discontinuity" around the value $30 \gev$. However, we traced the origin of this un-smoothness to our method of looping in order to compute $T_f$. Upon comparing with other packages (micrOMEGAs) around this point, the smoothness appears, and so the discontinuity is due to numerics and is not physical.

In this work focusing on model building, we carried out all the analytical and numerical analysis using Mathematica packages writing down explicitly the used formulae.
In future analysis, more sophisticated DM-oriented programs
such as microOMEGAs will be used, and we are currently investigating
the case for a more extensive set of benchmark points
Nonetheless, we checked, in a preliminary scan, using micrOMEGAs that Mathematica and micrOMEGAs results are similar. In Fig. (\ref{fig7}), we show the micrOMEGAs plot for the ``black'' relic density constraint, which shows, as expected, less stiffness at the $Z$-threshold and more smoothness around the resonance. Moreover, the values of $\lambda$ satisfying the relic density constraint in the neighborhood of the $Z$-threshold get smaller in Fig. \ref{fig7} compared to Fig. \ref{fig6}. This is justified as micrOMEGAs takes into account the fact that upon increasing the $DM$ mass, annihiliation starts before we reach the threshold \cite{gondolo}.


\begin{figure}[hbtp]
\centering
\includegraphics[scale=1]{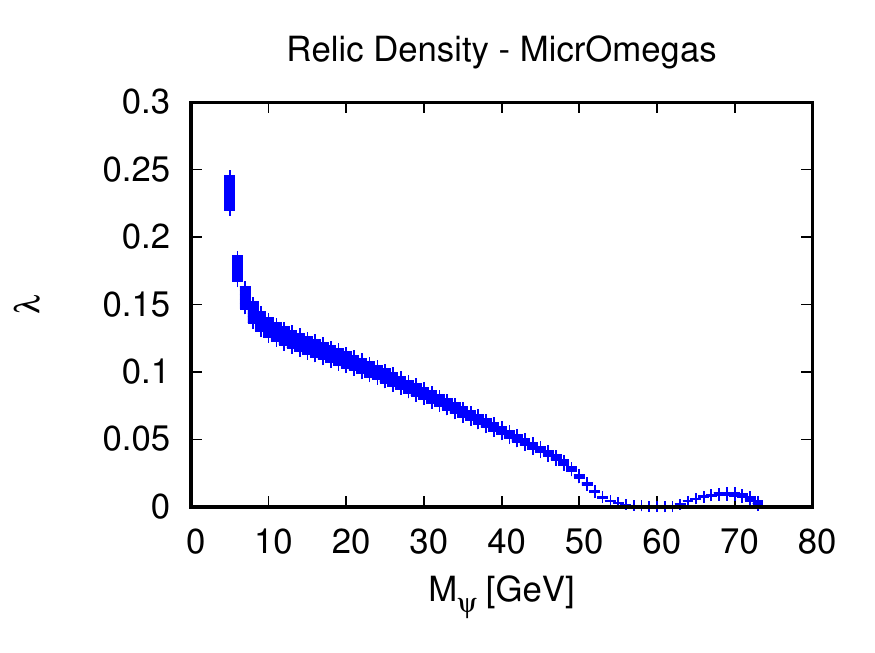}
\caption{{\footnotesize Relic density constraint using micrOMEGAs.}}
\label{fig8}
\end{figure}

It is noteworthy that the nonstandard interactions between the ``new inert'' scalars and the ``active'' Higgses bring about changes $\Delta S \, \& \,  \Delta T$ to the oblique electroweak parameters $S, T$ which encode the impact of new physics not coupled to SM fermions \cite{salah}. However, this electroweak precision testing goes beyond the scope of this work and will be dealt in the future.

In summary, we presented in this work a non-SUSY extension of the SM, where we
have six, instead of one, Higgs doublets. This helps to realize the
gauge coupling unification from one part. From the other part, the
lightest of the additional Higgs particles presents a suitable
candidate for DM. This has been illustrated for a specific benchmark point in the parameter space, and thus should stimulate
further investigation for other regions in the parameter space, a systematic full phenomenological analysis of which is a work in progress.

\section*{{\large \bf Acknowledgements}}
We thank Tianjun Li for useful discussions, and Salah Nasri for reading carefully the manuscript. N.C. wishes to
acknowledge support from the CAS PIFI scholarship, and the Humboldt Return fellowship. This
work is supported in part by the National Science Foundation of China and by the Chinese Academy of Sciences (the NSFC grant 11375248 for C.L. \& C.-S. H., and the NSFC grants 11373030 and 11633004 and CAS grant QYZDJ-SSW-SLH017 for X.C.).

\appendix

\section{2-loop RGE in 6HDM} \label{rge}
In order to derive  Eqs. (\ref{rge_g1^1}--\ref{rge_g3^2}), we note that the particle content of the 6HDM is just that of the SM plus ($n_h=5$) new Higgs doublets. Thus,  Eq. (\ref{beta}) shows that,  at 2-loop order, the 6HDM beta coefficients are just given by:
\bea
b_k = b_k^{\mbox{\tiny SM}} + b_k^{\mbox{\tiny new}} &,&  b_{k,\ell} = b_{k,\ell}^{\mbox{\tiny SM}} + b_{k,\ell}^{\mbox{\tiny new}}
\eea
The SM corresponding quantities are known \cite{rge2loop}:
\bea &
b_i^{\mbox{\tiny SM}}=\left(\begin{array}{c} 41/10 \\ -19/6 \\ -7 \end{array}\right), b_{ij}^{\mbox{\tiny SM}} = \left(\begin{array}{ccc} 199/50 & 27/10 & 44/5 \\ 9/10 & 35/6 & 12 \\ 11/10 & 9/2 & -26 \end{array}\right) ,& \\ & C^u_{\mbox{\tiny SM}}=\left(\begin{array}{c} -17/10 \\ -3/2 \\ -2 \end{array}\right), C^d_{\mbox{\tiny SM}}=\left(\begin{array}{c} -1/2 \\ -3/2 \\ -2 \end{array}\right),
C^e_{\mbox{\tiny SM}}=\left(\begin{array}{c} -3/2 \\ -1/2 \\ 0 \end{array}\right) &
\eea
We follow \cite{rgenew} to compute the contribution of new physics, and we get:
\bea
b_k^{\mbox{\tiny new}}=\left(\begin{array}{c} 1/10 \\ 1/6 \\ 0 \end{array}\right). n_h &,&  b_{k\ell}^{\mbox{\tiny new}} = \left(\begin{array}{ccc} 9/50 & 9/10 & 0 \\ 3/10 & 13/6 & 0 \\ 0 & 0 & 0 \end{array}\right). n_h
\eea
Replacing $n_h=5$ we get the desired equations. We checked using SARAH that we get the same 2-loop running beta functions for the 6HDM.

\section{Mass Eigenstates}

\subsection{Tadpole Equations} \label{Tadpole equations}
We get a local minima of $V_{\mbox{\tiny Active}}$ at $\phi_u=v_u, \phi_v=v_d$ when
\begin{align}
\m_{uu}  &= - \frac{1}{2} \lambda_{uuuu} v_{u}^{2}  - \frac{1}{2} v_{d}^{2} \Big(\eta_{uddu} + \lambda_{uudd} + \xi_{udud}\Big)\\
\m_{dd}  &= - \frac{1}{2} \lambda_{dddd} v_{d}^{2}  - \frac{1}{2} v_{u}^{2} \Big(\eta_{uddu} + \lambda_{uudd} + \xi_{udud}\Big)
\end{align}

\subsection{Mass Matrices}
\subsubsection{Mass Matrices for Scalars}
\label{Mass Matrices for Scalars}
\begin{enumerate}
\item {\bf Mass matrix for Higgs \label{Mass matrix for Higgs}}, Basis: \( \left(\phi_{u}, \phi_{d}\right), \left(\phi_{u}, \phi_{d}\right) \)
\begin{equation}
m^2_{h} = \left(
\begin{array}{cc}
\lambda_{uuuu} v_{u}^{2} &v_d v_u \Big(\eta_{uddu} + \lambda_{uudd} + \xi_{udud}\Big)\\
v_d v_u \Big(\eta_{uddu} + \lambda_{uudd} + \xi_{udud}\Big) &\lambda_{dddd} v_{d}^{2}
\end{array}
\right)
 \end{equation}
 The mass spectrum will read: \bea \label{m2evenspectrum}
M^2_{h_1,h_2}&=&\frac{1}{4}\left[(m^2_{h})_{11}+(m^2_{h})_{22}\pm
\sqrt{((m^2_{h})_{11}-(m^2_{h})_{22})^2+4((m^2_{h})_{12})^2}\right]\eea
This matrix is diagonalized, with mass eigenstates $(h_1,h_2)$, by \(Z^H = \left( \begin{array}{cc}
\cos \a &-\sin \a\\
\sin \a & \cos \a
\end{array}\right)\)
with \bea \sin 2\a =
\frac{2(m^2_{h})_{12}}{\sqrt{((m^2_{h})_{11}-(m^2_{h})_{22})^2+4(m^2_{h})^2_{12}}}\\\cos 2\a =
\frac{(m^2_{h})_{11}-(m^2_{h})_{22}}{\sqrt{((m^2_{h})_{11}-(m^2_{h})_{22})^2+4(m^2_{h})^2_{12}}}\eea

\item {\bf Mass matrix for Pseudo-Scalar Higgs \label{Mass matrix for Pseudo Higgs}}, Basis: \( \left(\sigma_{u}, \sigma_{d}\right), \left(\sigma_{u}, \sigma_{d}\right) \)
\begin{equation}
m^2_{A^0} = - \xi_{udud} \left(
\begin{array}{cc}
v_{d}^{2} &- v_d v_u\\
- v_d v_u &v_{u}^{2}
\end{array}
\right)
 \end{equation}
with squared mass of the non-Goldstone particle given by \bea \label{m2oddspectrum} M^2_o &=& - \xi_{udud} (v_d^2+v_u^2)\eea
This matrix is diagonalized, with mass eigenstates $(A^0_1,A^0_2)$, by \(Z^A=\left( \begin{array}{cc}
\cos \b &-\sin \b\\
\sin \b & \cos \b
\end{array}\right)\)

\item {\bf Mass matrix for Charged Higgs \label{Mass matrix for Charged Higgs}}, Basis: \( \left(H_u^{+,*}, H_d^-\right), \left(H_u^+, H_d^{-,*}\right) \)
\begin{equation}
m^2_{H^-} = - \frac{1}{2} \Big(\eta_{uddu} + \xi_{udud}\Big) \left(
\begin{array}{cc}
v_{d}^{2} &v_d v_u \\
v_d v_u &v_{u}^{2}
\end{array}
\right)
 \end{equation}
 with squared mass of the non-Goldstone particle given by \bea \label{m2chargedspectrum} M^2_c &=& -\frac{1}{2} \left(\xi_{udud}+\eta_{uddu}\right) (v_d^2+v_u^2)\eea

This matrix is diagonalized, with mass eigenstates $(H_1^{\pm},H_2^{\pm})$, by \(Z^+=Z^A\)

\item {\bf Mass matrix for Inert Higgs \label{Mass matrix for Inert Higgs}}, Basis: \( \left(\phi_{1}, \phi_{2}, \phi_{3}, \phi_{4}\right), \left(\phi_{1}, \phi_{2}, \phi_{3}, \phi_{4}\right) \)
\begin{align} \label{mphiphi} \left(m^2_{h^i}\right)_{\a\b} \equiv
m_{\phi_{\a}\phi_{\b}} &= \m_{\a \b } + \frac{1}{2} \sum_{I = u,d} {v_{I}^{2} \Big(\lambda_{II\a \b } + \eta_{I\a \b I} + \xi_{I\a I\b }\Big)}
\end{align}

\item {\bf Mass matrix for Inert Pseudo-Scalar Higgs \label{Mass matrix for Inert Pseudo Higgs}}, Basis: \( \left(\sigma_{1}, \sigma_{2}, \sigma_{3}, \sigma_{4}\right), \left(\sigma_{1}, \sigma_{2}, \sigma_{3}, \sigma_{4}\right) \)
\begin{align} \label{sigmasigma} \left(m^2_{A^{i0}} \right)_{\a\b} \equiv
m_{\sigma_{\a}\sigma_{\b}} &= \m_{\a \b} + \frac{1}{2} \sum_{I = u,d} v_{I}^{2} \Big(- \xi_{I\a I\b}  + \eta_{I\a \b I} + \lambda_{II\a \b}\Big)
\end{align}

\item {\bf Mass matrix for Inert Charged Higgs \label{Mass matrix for Inert Charged Higgs}}, Basis: \( \left(H_1^+, H_2^+, H_3^+, H_4^+\right), \left(H_1^{+,*}, H_2^{+,*}, H_3^{+,*}, H_4^{+,*}\right) \)
\begin{align} \label{h+h+} \left(m^2_{H^{i+}} \right)_{\a\b} \equiv
m_{H_\a ^+H_\b ^{+,*}} &= \m_{\a \b} + \frac{1}{2} \sum_{I = u,d} \lambda_{II\a \b} v_{I}^{2}
\end{align}

\end{enumerate}

\subsection{Scalar Particle Contents}
\label{Particle Contents}
\begin{center}
\begin{longtable}{l||c|c|c}
\hline \hline
Name  & complex/real & Generations & at ``symmetric" benchmark point ($\b=45^o=-\a,M_{h_1}=M_{h_2}$)\\
\hline \hline
\(h\)  &real&2&\\\hline
 \(A^0\) &real&2& {\(A^0_1 \rightarrow \)  goldstone boson, \(A^0_2 \rightarrow \) heavy}\\\hline
 \(H^-\)  &complex&2& {\(H^-_1 \rightarrow \) goldstone boson, \(H^-_2 \rightarrow \) heavy}\\\hline
 \(h^i\) &real&4& {\(h^i_1 \rightarrow \) Dark Matter Candidate $\equiv \psi$,}\\
 &&&{\(h^i_\a \rightarrow \) heavy where \(\a = 2,3,4 \)}\\\hline
  \(A^{i0}\)  &real&4&{\(A^{i0}_\a \rightarrow \) heavy where \(\a = 1,2,3,4 \)}\\\hline
 \(H^{i+}\)  &complex&4&{\(H^{i+}_\a \rightarrow \) heavy where \(\a = 1,2,3,4 \)}\\
  \hline \hline

\end{longtable}
 \end{center}

\section{Annihilation amplitudes} \label{annihilation}
\begin{itemize}
\item $\psi \psi \rightarrow W^+ W^-$ Channel:
\begin{dmath} \label{M2ww-channel}
|\mathcal{M}|^2=\frac{\left(-4 S M_W^2+12 M_W^4+S^2\right) \left(-g_2^2 M_H^2+g_2^2 S+4 \lambda  M_W^2\right){}^2}{16 M_W^4 \left(S-M_H^2\right){}^2}
\end{dmath}
\hrule

\item $\psi \psi \rightarrow Z Z$ Channel:
\begin{dmath} \label{M2zz-channel}|\mathcal{M}|^2=
\frac{\left(-4 S M_Z^2+12 M_Z^4+S^2\right) \sec ^4\left(\theta _W\right) \left(-g_2^2 M_H^2+g_2^2 S+4 \lambda  M_W^2\right){}^2}{16 M_Z^4 \left(S-M_H^2\right)^2}
\end{dmath}
\hrule

\item $\psi \psi \rightarrow H H$ Channel:
\begin{dmath}\label{M12HH-channel}|\mathcal{M}_{h_1h_1}|^2=
\left(\lambda -\frac{12 \lambda  M_H^2 M_W^2}{g_2^2 \text{v}^2 \left(M_H^2-S\right)}\right){}^2
\end{dmath}
\begin{dmath} \label{M22HH-channel}|\mathcal{M}_{h_2h_2}|^2=
\lambda ^2 \left(\frac{4 M_W^2 \left(\frac{3 M_H^2}{\text{v}^2 \left(S-M_H^2\right)}+\lambda  \left(\frac{1}{T-M_{\psi }^2}+\frac{1}{U-M_{\psi }^2}\right)\right)}{g_2^2}+1\right){}^2
\end{dmath}
and we get upon integration over the phase space:
\begin{dmath}\label{wHH-channel}\omega(S)=
\frac{16 \lambda ^2 M_W^2}{g_2^4} \left(\frac{\lambda  \log\left( \frac{S-2 M_H^2-\sqrt{S-4 M_H^2}\sqrt{S-4 M_{\psi}^2}}{S-2 M_H^2+\sqrt{S-4 M_H^2}\sqrt{S-4 M_{\psi}^2}}\right) M_H^2 \left(4 M_W^2 \left(3 S+\lambda  v^2\right)-3 g_2^2 S v^2\right)}{v^2 \sqrt{S-4 M_H^2} \left(-3 S M_H^2+2 M_H^4+S^2\right) \sqrt{S-4 M_{\psi }^2}}+\frac{2 \lambda  \log\left( \frac{S-2 M_H^2-\sqrt{S-4 M_H^2}\sqrt{S-4 M_{\psi}^2}}{S-2 M_H^2+\sqrt{S-4 M_H^2}\sqrt{S-4 M_{\psi}^2}}\right) M_H^4 \left(g_2^2 v^2-12 M_W^2\right)}{v^2 \sqrt{S-4 M_H^2} \left(-3 S M_H^2+2 M_H^4+S^2\right) \sqrt{S-4 M_{\psi }^2}}+\frac{\lambda  \log\left( \frac{S-2 M_H^2-\sqrt{S-4 M_H^2}\sqrt{S-4 M_{\psi}^2}}{S-2 M_H^2+\sqrt{S-4 M_H^2}\sqrt{S-4 M_{\psi}^2}}\right) S v^2 \left(g_2^2 S-4 \lambda  M_W^2\right)}{v^2 \sqrt{S-4 M_H^2} \left(-3 S M_H^2+2 M_H^4+S^2\right) \sqrt{S-4 M_{\psi }^2}}+\frac{3 M_H^2 \left(M_H^2 \left(6 M_W^2-g_2^2 v^2\right)+g_2^2 S v^2\right)}{v^4 \left(S-M_H^2\right){}^2}+\frac{2 \lambda ^2 M_W^2}{-4 M_H^2 M_{\psi }^2+M_H^4+S M_{\psi }^2}+\frac{1}{8 M_W^2}\right)
\end{dmath}
\hrule

\item $\psi \psi \rightarrow e^+ e^-$ Channel:
\begin{dmath}\label{M2ee-channel}|\mathcal{M}|^2=
\frac{2 \lambda ^2 M_e^2 \left(S-4 M_e^2\right)}{\left(S-M_H^2\right){}^2}
\end{dmath}
\hrule

\item $\psi \psi \rightarrow u \bar{u}$ Channel:
\begin{dmath} \label{M2uu-channel}|\mathcal{M}|^2=
\frac{6 \lambda ^2 M_u^2 \left(S-4 M_u^2\right)}{\left(S-M_H^2\right){}^2}
\end{dmath}
\hrule

\item $\psi \psi \rightarrow d \bar{d}$ Channel:
\begin{dmath}\label{M2dd-channel}|\mathcal{M}|^2=
\frac{6 \lambda ^2 M_d^2 \left(S-4 M_d^2\right)}{\left(S-M_H^2\right){}^2}
\end{dmath}
\hrule

\end{itemize}

\section{Direct Detection Amplitudes} \label{DDamplitudes}

\begin{dmath}\label{M2DD}|\mathcal{M}|^2=
-\frac{3 \lambda ^2 M_{\text{N}}^2 \eta _{\text{N}}^2 \left(T-4 M_{\text{N}}^2 \right)}{\left(T-M_H^2\right){}^2}
\end{dmath}

The thermal average for total scattering:
\begin{dmath} \label{sigmavDD}\langle \sigma v_{rel}\rangle=
\frac{12 \lambda ^2 M_{\text{N}}^4 \eta _{\text{N}}^2}{M_H^4}+\frac{6 \lambda ^2 p_f^2 M_{\text{N}}^2 \eta _{\text{N}}^2 \left(M_H^2-8 M_{\text{N}}^2 \right)}{M_H^6}+O\left(p_f^3\right)
\end{dmath}

\section{Invisible Higgs Decay} \label{IHDappendix}

\begin{itemize}

\item $h_i \rightarrow f\bar{f}$
\begin{dmath} \label{fM2IHD}|\mathcal{M}|^2=
\frac{C_f g_2^2 M_f^2 \left(M_H^2-4 M_f^2\right)}{4 M_W^2}
\end{dmath}
where:\\
$f = u,d,e$ \& $i = 1,2$\\
$C_f = 1$ for leptons\\
$C_f = 3$ for quarks\\
\hrule

\item $h_2 \rightarrow ZZ$
\begin{dmath} \label{ZM2IHD}|\mathcal{M}|^2=
\frac{g_2^2 M_W^2 \left(-4 M_H^2 M_Z^2+M_H^4+12 M_Z^4\right) \sec ^4\left(\theta _W\right)}{4 M_Z^4}
\end{dmath}
\hrule

\item $h_2 \rightarrow W^+W^-$
\begin{dmath} \label{WM2IHD}|\mathcal{M}|^2=
\frac{g_2^2 \left(-4 M_H^2 M_W^2+M_H^4+12 M_W^4\right)}{4 M_W^2}
\end{dmath}
\hrule

\item $h_2 \rightarrow \psi_k\psi_j$
\begin{dmath} \label{psiM2IHD}|\mathcal{M}^2|=
\frac{4 \lambda ^2 M_W^2}{g_2^2}
\end{dmath}
\hrule

\end{itemize}

\section{self Elastic Scattering Amplitudes} \label{self elastic scattering amplitudes}

\begin{dmath} \label{M2SES}|\mathcal{M}|^2=
\left(\frac{4 \lambda ^2 M_W^2 \left(\frac{1}{S-M_H^2}+\frac{1}{T-M_H^2}+\frac{1}{U-M_H^2}\right)}{g_2^2}+\lambda _S\right){}^2
\end{dmath}

The total scattering:
\begin{dmath} \label{sigmavSES} \langle \sigma_{el}\rangle=
\frac{\left(g_2^2 M_H^4 \lambda _s-4 M_H^2 \left(g_2^2 M_{\psi }^2 \lambda _s+3 \lambda ^2 M_W^2\right)+32 \lambda ^2 M_W^2 M_{\psi }^2\right){}^2}{g_2^4 M_H^4 \left(M_H^2-4 M_{\psi }^2\right){}^2}-\frac{256 p_f^2 \left(\lambda ^2 M_W^2 M_{\psi }^2 \left(M_H^2-2 M_{\psi }^2\right) \left(g_2^2 M_H^4 \lambda _s-4 M_H^2 \left(g_2^2 M_{\psi }^2 \lambda _s+3 \lambda ^2 M_W^2\right)+32 \lambda ^2 M_W^2 M_{\psi }^2\right)\right)}{g_2^4 M_H^6 \left(M_H^2-4 M_{\psi }^2\right){}^3}+O\left(p_f^3\right)
\end{dmath}

\end{document}